\documentclass[11pt]{article}

\batchmode

\usepackage{amsfonts}

\newtheorem{theorem}{Theorem}
\newtheorem{lemma}{Lemma}

\newtheorem{definition}{Definition}

\def\lor{\mathop{\mbox{\Large $($}}}  
\def\ler{\mathop{\mbox{\Large $)$}}}  
\def\geqq{\mathop{\;\geq \;}}  
\def\leqq{\mathop{\;\leq \;}}  
\def\={\mathop{\;=\;}}  
\def\largabs{\mathop{\mbox{\large $|$}}}

\def\cali{\mathop{\cal I}}

\def\uin{\mathop{\;\underline\in\;}}
\def\vareps{\mathop{\varepsilon}}

\newcounter{aid}

\begin{document}

\title{\bf  
Almost Uniform Sampling of Independent Sets in Polynomial Time -- Implying {\bf NP}={\bf RP}}

\author{ \vspace*{1mm} Andr\'as  Farag\'o \\
Department of Computer  Science      \\
The  University   of  Texas   at  Dallas\\
   Richardson,  Texas \\
   {\tt farago@utdallas.edu} }

\date{}
\maketitle

\begin{abstract} {\bf\em We prove the unexpected result
that almost uniform sampling of independent sets in graphs is possible via a 
probabilistic polynomial time algorithm.  Note that our sampling algorithm (if correct) has {\em extremely surprising} consequences; 
the most  important one being no less than the unlikely collapse {\bf NP}={\bf RP}.

}
 \end{abstract}

\thispagestyle{empty}

\newpage
\setcounter{page}{1}

\section{Introduction}

It has been well known for decades that the approximately  uniform
random  sampling  of  combinatorial  structures  and  the  approximate
counting of these  structures are closely  related, see e.g.,  Jerrum,
Valiant,   and   Vazirani   \cite{jerrum2},   Sinclair   and    Jerrum
\cite{sinclair2}. Moreover, it is also known that  efficient
algorithms for such tasks, if exist, can imply a dramatic collapse of
complexity classes.

Regarding random sampling,  generating
a  uniform   random  sample  from  a family that is
exponentially large  in terms  of its  defining description  is a hard
problem for many  such sets. A  typical example is producing a uniform
random independent set from a graph.  It is hard in the following  sense
(Sinclair \cite{sinclair}):  if we can  generate a {\em perfectly uniform}  random
sample  from  the  independent  vertex  sets  of any graph in
probabilistic  polynomial time, then ${\bf NP}={\bf RP}$.

The above hardness result is not too difficult to prove (see Sinclair \cite{sinclair}, Theorem 1.17), but it leads to other,
more challenging questions. What happens if the graph is restricted to a special class, for example, graphs with 
a bounded maximum degree? 
What if we can generate a random sample that is only approximately (but not perfectly) 
uniform? How good approximation can be achieved efficiently, and in which classes of graphs? 
In many cases it is known that the task still remains hard (see, e.g., Dyer, Frieze and
Jerrum \cite{dyer}),  but the analysis is much more involved.

The standard and well studied approach to random sampling problems  is
the Markov  Chain Monte  Carlo (MCMC)  method; for  surveys see, e.g., 
Jerrum and Sinclair \cite{jerrum1},
Randall  \cite{randall}, Levin,
Peres, and Wilmer \cite{levin}, Mikl\'os \cite{miklos}, 
Brooks, Gelman, Jones, and  Meng (Eds.) \cite{brooks}, and many others.
The method works by designing an  appropriate
Markov chain on the  set from which we  want to sample, such  that the
stationary  distribution  of  the  chain  is some desired distribution
(typically uniform) and  the chain is  defined by simple  local rules,
which  is  usually  easily  achievable.  Then,  if the Markov chain is
mixing fast enough, i.e., it approaches the stationary distribution rapidly 
(which may be challenging to prove), we can  obtain
an  almost  uniform  random  sample  by  running it for a polynomially
bounded time.

Building  on  this  approach  has  resulted  in notable successes that
provided polynomial-time randomized approximation schemes for a number
of  important  problems,  such  as  approximating the volume of convex
bodies (Dyer,  Frieze and  Kannan \cite{dyerfrieze},  Kannan, Lov\'asz
and    Simonovits    \cite{kannan,lovasz_simonovits}),   approximately
counting  perfect  matchings   or,  equivalently,  approximating   the
permanent of a 0-1 matrix (Jerrum and Sinclair \cite{jerrumsinclair}),
and a good number of others. In hard cases (the ones that would yield ${\bf
NP}={\bf RP}$), however, this approach runs into the severe obstacle of  {\em slow
mixing,} that is, the Markov chain may require exponentially long  time
to get close enough to its stationary distribution. Specifically, this
is the case  for independent vertex  sets in graphs.  Dyer, Frieze and
Jerrum \cite{dyer} proved (unconditionally) that in graphs with maximum degree at  least
6, {\em any}  Markov chain  on the  independent sets  has exponential mixing
time  in  the  worst  case,  given  that  the  chain  is  defined with
sufficient  locality\footnote{Specifically, what  they  require  is  that the Markov
chain changes the status of at most 35\% of vertices in a single step.
This is satisfied by  all known Markov chains  that are used for  this
problem.}.

It is worth  noting that rather  than only approaching  the stationary
distribution of  a Markov  chain, it  is possible  to {\em  perfectly}
reach  it within finite time.  That  is,  there  are  algorithms that can generate a {\em
precisely}  stationary  sample  (without  knowing  or  computing   the
distribution). Examples of such {\em perfect sampling} algorithms  are
the Coupling From  the Past method  of Propp and  Wilson \cite{propp},
Fill's  algorithm  \cite{fill}  and  the  exact  mixing  algorithm  of
Lov\'asz and Winkler  \cite{lovasz}. Unfortunately, however,  for such
slowly mixing Markov chains, as the ones used for independent sets  in
graphs, all the known  perfect sampling methods need  exponential time
in the worst case.

The  task  of  (near-)perfect  sampling  is  a  key  ingredient in our
solution, but we give a significant  new twist to it. We ``shift"  the
hardness by  assuming that  the Markov  chain is  {\em already  in the
stationary distribution}. Of course, without any other condition, this
would make  the sampling  task trivial.  But we  also require that the
output  sample  must  be  restricted  to  a given {\em
subset} of the  state space that  we call {\em  target set.} Thus,  in
this task of {\em Subset Sampling}, we want to transform a  stationary
sample into another one that  is conditioned on falling in  the target
set.  This  shifts  the  hardness  in  the  sense that the hardness of
reaching the stationary distribution is  already incorporated in  the input, we  only
want to somehow ``corner it" into the target set.

Let us mention that if we  could {\em perfectly}  accomplish subset  sampling for  slowly mixing
chains, then it would allow an easy recursive algorithm for efficient random
sampling from the independent vertex  sets of  a graph, which would yield  ${\bf NP}={\bf RP}$ (see Sinclair \cite{sinclair}).
 One could  start with an edgeless
graph, where the task is trivial, and gradually restrict the family of
allowed independent sets to smaller and smaller subfamilies, by adding
the edges  one by  one, until  the desired  graph is reached. If we could
always {\em perfectly} accomplish the ``cornering" of the stationary  sample
to a target set, then a direct efficient solution would result. Unfortunately, we are 
not that lucky, as we can only approximate the perfect subset sample, 
which requires significant effort to keep the accumulating errors under control.

Note  that  the  naive  approach  to  Subset Sampling ``run the chain,
starting from  the stationary  input sample,  until the  target set is
hit" may create a strongly biased output. A simple example is a random
walk on a graph that is a large circle, and the target set consists of
a path of 3  consecutive vertices along the  circle. Let us start  the
walk from a  vertex chosen uniformly  at random. If  the initial state
does not happen to  fall in the 3-node  path (which is very likely if
the circle is large), then we can only hit the target set first at one
of its end-vertices. As a result, the first hitting place will be  far
from uniformly distributed on the target set, despite to starting  the
chain from its uniform stationary distribution.

A natural next approach to  try could be designing more  sophisticated
{\em stopping  rules}, rather  than just  the first  hit of the target
set. For  example, we  could say  ``run the  chain from the stationary
initial state until the empirical  distribution of hits in the  target
set  is  close  enough  to  the  conditioned stationary distribution."
Unfortunately, however, no stopping rule is known that would carry out Subset Sampling 
efficiently in general for slowly mixing Markov chains. For results about stopping rules
see, e.g., Lov\'asz and Winkler \cite{lovasz_stoc,lovaszmixing}.

In view of all the negative results regarding hard sampling  problems,
and considering the unlikely consequences, it seems rather hopeless to
efficiently  accomplish  Subset  Sampling  perfectly, and in full
generality. Nevertheless, it
turns  out,  surprisingly  enough,  that  if  we  are satisfied with a
near-perfect subset  sample, and with a special target set, then  it
is still solvable efficiently,
even  in  case  of  a  slowly  mixing  chain, on a state space that is
exponentially large in terms of its description. This will be
enough to allow us to prove ${\bf NP}={\bf RP}$.

Quite interestingly, however, we cannot  achieve Subset Sampling with a single sample alone. Our key enabling result is 
that if we have (polynomially) many {\em independent} realizations of the Markov chain, then we can generate  
equally many {\em independent} subset samples from them, in probabilistic  polynomial time. 
We call this task {\em Collective Subset Sampling}. Thus, curiously, we can collectively subset sample independent realizations efficiently, 
yielding independent subset samples, one  for each realization, but we do not know how to obtain a {\em single} 
subset sample efficiently from a {\em single} realization. In this sense, the multitude of independent samples carries useful extra information, 
which appears to be necessary to achieve the goal.

Since we  solve Subset  Sampling only  approximately, a  great deal of
effort has to go into  estimating the errors, and their  accumulation.
To this end,  a distance measure  of probability distributions  called
{\em separation distance} will play an important role, as some of  its
properties,  which we  prove  in  Lemma~\ref{lemma1}, are very 
helpful.

As a final introductory note, we do not claim that our proof is the simplest possible, and the algorithm is the fastest possible.
Rather than optimizing the algorithm, our goal is to present all new proofs in  more  details  than  what  is  customary in most papers, to 
make the checking of correctness as feasible as possible, to make sure that the main claim ({\bf NP}={\bf RP})
indeed holds.  

\section{Roadmap to the Results}

\begin{itemize}

\item 
After summarizing a few preliminaries in Section~\ref{prelim}, in Section~\ref{tools} we introduce some useful probabilistic tools. In particular we discuss 
the separation distance of probability distributions in Section~\ref{sepdist}, and a Markov chain on the independent sets of a graph in Section~\ref{BIDC}.

\item 
In Section~\ref{subsetsamp} we introduce and analyze a key tool that we call {\em Independence Property}. The essence of this property is 
that if independent Markov chain realizations are filled into the rows of a square matrix, then we can select an entry from each row, such that each 
follows the stationary distribution conditioned on the target set, and they are  also {\em independent}. A central tool to achieve this is using a 
uniform random perfect matching in a bipartite graph. In Section~\ref{subsetsamp} we analyze the ideal case, when the random perfect matching is
perfectly uniform. Later, in Sections~\ref{HPM} and \ref{analysis}, we are going to approximate it by a probabilistic  polynomial time  algorithm of Jerrum, Sinclair and Vigoda \cite{jerrum}, still keeping the arising deviation under control.

\item In Section~\ref{almost} we introduce the actual sampling algorithm. The algorithm uses two sub-routines, which are discussed in 
Section~\ref{HPM}. One of them decides whether a square matrix has a perfect matching (as a generalization of the bipartite graph matching).
If it exists, then the other subroutine generates it approximately uniformly at random, building on the above mentioned result of 
Jerrum, Sinclair and Vigoda \cite{jerrum}.

\item In Section~\ref{analysis} we analyze the main algorithm, and prove that it indeed delivers an  approximately uniform  independent set efficiently, 
as promised. A part of the proof is the analysis of the probability that the used random square matrix indeed has a perfect matching (as generalized for matrices). This is somewhat more complex to prove than one would expect, due to the non-independence of the matrix entries. Therefore, we present
this proof in a separate section (Section~\ref{probpm}) along with an appendix (Appendix B).

\item In Section~\ref{FPRAS} we show how to use our sampling algorithm to derive a Fully Polynomial Approximation Scheme (FPRAS) for 
counting independent sets in bounded degree graphs.

\item Section~\ref{NPvsRP} we use known results to derive ${\bf NP}={\bf RP}$ from the existence of the aforementioned FPRAS.

\item Finally, in Section~\ref{other} we discuss some consequences of the main result. One such consequence is that the entire Polynomial Hierarchy 
collapses into {\bf BPP}, i.e., {\bf PH}={\bf BPP}. Another consequence is that the polynomial time  solvability of {\bf NP}-complete problems 
{\em on the average} now becomes equivalent to their polynomial time  solvability in the {\em worst case.} These were long standing open problems.

\end{itemize}

\section{Preliminaries}
\label{prelim}

The ranges of random variables are always finite in this  paper,
so we do not repeat this with each statement and definition.
The probability distribution of a random variable $\xi$ with range $S$
is  represented  by  a  finite  dimensional  vector. For such a vector
$\alpha$,  its  components  are  $\alpha(x)=\Pr(\xi=x)$, as $x$ ranges
over $S$ in some order. The actual ordering will be irrelevant, so  we
just assume any fixed  ordering. The inequality $\alpha\geq \beta$ means 
$\alpha(x)\geq \beta(x)$ for every $x\in S$. If $H\subseteq S$, then 
$\alpha(H)=\sum_{x\in H}\alpha(x)$.

For graphs the standard notation $G=(V,E)$ is used, where $V=V(G)$ is the set of
vertices and $E=E(G)$ is the set of edges. 
For a bipartite graph, its {\em bi-adjacency 
matrix} is a  matrix $[a_{ij}]$ in which each row represents a vertex on one side 
of the bipartition, say, left side, each column stands for a vertex on the right side, and 
$a_{ij}=1$ if left-side vertex $v_i$ is connected by an edge to right-side vertex $v_j$, otherwise 
$a_{ij}=0$. 
The bipartite graph is called {\em balanced} if its bi-adjacency matrix is a square matrix, i.e., 
there is an equal number of vertices on the two sides of the graph. 
If $B$ is a square 0--1 matrix, then the bipartite graph with  bi-adjacency matrix $B$ is denoted by $G_B$.

Another notation that we often apply: 
if $n$ is a positive integer, then $[n]$ denotes the set $\{1,\ldots,n\}$.



\section{Probabilistic Tools}
\label{tools}

\subsection{ Separation Distance of Probability Distributions}
\label{sepdist}

In approximating probability distributions  it is important to measure 
how different two distributions are.
There are many ways to  measure the the deviation between  probability
distributions; for  a comprehensive  survey see,  e.g., Deza  and Deza
\cite{deza}. For  our purposes  the concept  known as  {\em separation
distance}  will  serve the  best. It was
used by Aldous and Diaconis \cite{aldous-diaconis} in the  analysis of  Markov
chains; we define it here slightly differently.

\begin{definition} \label{distance}
{\bf (Separation distance)}
 For  any two probability distributions $\alpha, \beta$ on the same
 set $S$,
 their {\em separation distance}, denoted by
 $s(\alpha,\beta),$ is defined as
\begin{equation}\label{def1formula}
 s(\alpha,\beta)=  \min\{\varepsilon\geq 0\;|\;
 \alpha \geq (1-\varepsilon)\beta \}.
 \end{equation}
 \end{definition}
 
\noindent 
For example, $s(\alpha,\beta)=0.01$ means that for every $x\in S$ we have $\alpha(x)\geq 0.99\beta(x)$, but for any $c>0.99$ there is an $x\in S$ with 
$\alpha(x)< c\beta(x)$.


\noindent {\em Remark:}
 Aldous and Diaconis \cite{aldous-diaconis} defined the same distance concept by
 \begin{equation}\label{ald}
 s(\alpha,\beta)=\max_x \left(1-\frac{\alpha(x)}{\beta(x)}\right).
 \end{equation}
 It is easy to see that
 the two definitions are equivalent, assuming that in (\ref{ald}) we
take care of the case
$\beta(x)=0$. 


Let us mention that the separation distance is not a metric in the
usual sense, since it is not symmetric\footnote{A symmetric version
was defined by Sinclair and Jerrum \cite{sinclair2}, under the name
{\em relative pointwise distance}. 
We stay here with the non-symmetric version, as its properties better fit our purposes.}. 
It can also significantly differ from the well known total
variation distance. 
In fact, $s(\alpha,\beta)$ is always an upper bound on the total variation distance, see (v) in Lemma~\ref{lemma1} below.

In the next lemma we summarize some useful properties of the separation distance.
\vspace*{-3mm}
\begin{lemma} \label{lemma1}
For arbitrary probability distributions on the same finite
set $S$, their separation distance has the following
properties:
\begin{itemize}\itemsep1mm
\item[\rm (i)]
{\bf\em Normalized Value:}
$0\leq s(\alpha,\beta)\leq 1$.
\item[\rm (ii)]
{\bf\em Self-Identity:}
$s(\alpha,\beta)=0$ holds if and only if $\alpha=\beta$.
\item[\rm (iii)]
{\bf\em Triangle Inequality:} $s(\alpha,\beta)+s(\beta,
\gamma) \geq s(\alpha, \gamma)$.
\item[\rm (iv)]
{\bf\em Mixture Representation:}
For any $0\leq  \varepsilon\leq 1$, the inequality   $s(\alpha,\beta)\leq \varepsilon$
holds if and only if there exists a probability distribution $\gamma$ on $S$, satisfying
$$\alpha = (1-\varepsilon) \beta + \varepsilon \gamma.$$

\item[\rm (v)]
{\bf\em Connection to total variation distance:} The {\em total variation distance} of $\alpha, \beta$ is 
$$
d_{TV}(\alpha, \beta) \= \frac{1}{2}\sum_{x\in S} |\alpha(x)-\beta(x)| \= \max_{A\subseteq S} |\alpha(A)-\beta(A)|.
$$
The separation distance is always an upper bound for $d_{TV}$: 
$$d_{TV}(\alpha, \beta) \leqq s(\alpha, \beta).$$

\item[\rm (vi)]
{\bf\em Relationship to total variation distance for uniform distribution:} If $U$ is the  uniform distribution on $S$, then
$$
s(\alpha, U) \leqq 2|S|d_{TV}(\alpha, U).
$$


\end{itemize}
\end{lemma}

\noindent {\bf Proof:}
See Appendix A.

\subsection{A Markov Chain on the Independent Sets of a Graph}
\label{BIDC}

Let us introduce a Markov chain defined on the independent sets of  a graph $G=(V,E)$.
We  call it {\em Basic Insert/Delete Chain (BIDC)}. It is similar to what is called Insert/Delete Chain  by
Dyer  and  Greenhill  in  \cite{dyergreenhill}\footnote{Related models  are also frequently
investigated  in  the  statistical  physics  literature, under various names,  such  as  hard-core  lattice  gas  model,
Glauber dynamics, Potts model, Ising model, etc. Our intent, however, is to abstract away from the statistical physics 
background, and focus rather on the graph theoretic connection.}. We distinguish our version with the adjective ``basic", as
we do not need the additional parameter $\lambda$ (sometimes  referred
to  as  {\em  fugacity},  which   weighs  a  set  of  size   $k$  with
$k^\lambda$), and  also we  will not  need to  keep the  chain in  the
current state with probability $\geq 1/2$.

The transition probabilities of our Markov chain are given as follows.
Let ${\cal I}={\cal I}(G)$ denote the family of independent sets of $G$ (including the empty set).
Let $X_t\in\cal I$ be the state of the chain at time $t$. (The time is
discrete,  so  $t$  is  assumed  an  integer.)  Given the current state $X_t$, the  next  state
$X_{t+1}\in\cal I$ is obtained by the following rule.

\begin{quote}
\noindent {\bf Basic Insert/Delete Chain (BIDC) transition
rule}
\begin{enumerate}
\item
Draw a vertex $u\in V$ uniformly at random.
\item Set
\begin{eqnarray} \nonumber
X_{t+1}= \left\{
\begin{array}{cl}  X_t-  \{u\}  &  \;\;\mbox{  if  }\;\;\;u\in  X_t \\
X_t\cup \{u\} & \;\;\mbox{ if }\;\;\;u\notin X_t \mbox{ and }  X_t\cup
\{u\}\in\cali \\ X_t & \;\; \mbox{ otherwise. }
\end{array} \right.
\end{eqnarray}
\end{enumerate}
\end{quote}

{\bf Properties of BIDC.} The introduced  Markov chain can be interpreted as a random walk on an
 undirected graph $\widetilde G$, with $V(\widetilde G)={\cal I}(G)$. As each vertex of $\widetilde G$ represents an independent
set of  $G$,  therefore,  $\widetilde G$ is typically exponentially large in terms 
$|V|=n$. Two different vertices  $X,Y$
are connected by an edge in $\widetilde G$ if and
only if $|X\Delta Y|=1$, where $\Delta$ denotes symmetric difference.
There are also self-loops, making the graph regular of degree
$n$, since the random vertex $u$ can be chosen $n$ different ways (a self-loop contributes 1 to the degree).
This graph is  always connected,  since all independent sets can be
reached
from  $\emptyset$.  It  follows  from  standard results on Markov
chains and  random walks  on graphs  (see, e.g.,  the survey
by Lov\'asz \cite{lovasz2}) that this Markov chain has
a unique stationary distribution, which is
the uniform distribution on ${\cal I}(G)$.
On the other hand, the mixing time is exponentially large in $n$ in the worst case, 
whenever the graph has maximum degree at least 6, as implied by  
the results of Dyer,  Frieze  and Jerrum \cite{dyer}.

\section{A Key Tool for Collective Subset Sampling: The Independence Property}
\label{subsetsamp}

As already mentioned in the Introduction, a key tool to achieve our goals will be Subset Sampling. It means that having received an (almost)
stationary sample of a Markov chain as input, the output is another random sample that is confined to a subset of the state space, called {\em target set,}
such that its distribution preserves the (almost) stationary distribution, conditioned on the target set.
However, we do not know how to achieve Subset Sampling with a {\em single sample} alone. Our key enabling result is 
that if we have (polynomially) many {\em independent} realizations of the Markov chain, then we can generate  
equally many {\em independent} subset samples from them, in probabilistic  polynomial time. 
We call this task {\em Collective Subset Sampling}. 
In this sense, the multitude of independent samples carries useful extra information, 
which appears to be necessary to achieve the goal.

In Collective Subset Sampling we handle the samples such that an independent stationary realization of the considered Markov chain is filled into each row 
of a square matrix.  Then we pick certain matrix elements, reducing it to the problem of generating a uniform perfect matching in a bipartite graph.
In the present section, as the source of the approach, we still assume that an ideally uniform perfect matching can be generated. 

While we do not know how to actually generate an {\em ideally uniform} random perfect matching, 
later we are going to replace it by an algorithm of Jerrum, Sinclair and Vigoda \cite{jerrum} which constructs an {\em almost uniform} random 
perfect matching in a bipartite graph in probabilistic polynomial time. We will still be able to keep the arising deviation under control.

A central feature of the resulting sample system is that in the ideal case (when an ideally uniform random perfect matching is used), 
the obtained samples follow the stationary distribution conditioned on the target set, and they will be {\em independent.} 
We refer to this as the {\em Independence Property}, which plays a key role 
in enabling Collective Subset Sampling. In the present section we elaborate on the Independence Property in a somewhat more abstract setting.

\subsection{Notations and Notions}
\label{noti}

For an integer $n\geq 1$ let $X=[X_{ij}]$ be an $n\times n$-sized random matrix with the following properties:
\begin{description}

\item[\rm $\;\;\;$ (1)] 
 Each entry $X_{ij}$ is a random variable, taking its values from a common non-empty finite domain $S$, 
referred to as state space. 

\item[\rm $\;\;\;$ (2)]
The probability distributions of the entries $X_{ij}$ over $S$ are arbitrary and possibly different, except that they are required to satisfy  
the conditions given in items (3) and (4) below.

\item[\rm $\;\;\;$ (3)] 
For a fixed subset $H\subseteq S$ it holds that 
$\Pr(X_{ij} \in H)>0$ for every $i, j\in [n]$, and the conditional probability distribution of any entry $X_{ij}$, 
given $X_{ij}\in H$, is the same fixed distribution $\alpha$ on $H$ for each entry. 
Formally,
$$
\forall i, j\in [n] \;\; \forall x\in H : \;\;\;\; \Pr\lor X_{ij}=x \;|\;X_{ij} \in H \ler \= \alpha(x).
$$

\item[\rm $\;\;\;$ (4)]
The row vectors of $X$, as random vectors, are independent of each other. (On the other hand, no independence is required among the entries 
{\em within} any given row.)

\end{description}

\begin{definition}\label{conform} {\bf ($(H,\alpha)$-anchored random matrix)}
If a random matrix $X$ satisfies the above conditions {\rm (1), (2), (3), (4),} then $X$ is called an $(H,\alpha)$-anchored random matrix.
\end{definition}
{\bf Remark.} The square matrix $X$ will be constructed such that an independent stationary realization of the considered Markov chain is filled into each row. 
The subset $H$ is the target set of Subset Sampling. Property (3) expresses the fact that the stationary Markov chain remains stationary 
when conditioned on falling into the target set. Property (4) reflects that the realizations filled into different rows are independent, but the entries of the 
same realization (i.e., the entries in any single row) are not. These properties are stated somewhat more generally than the Markov chain setting,
as we only want to use features that are indeed needed.

Now we introduce the concept of a {\em perfect matching} in a matrix. The name comes from the analogy with a perfect matching in bipartite graphs, as explained later. All definitions in this section assume an $n\times n$-sized matrix $X$, in which each entry takes its values from a common finite domain $S$, so we do not repeat this condition in each definition.

\begin{definition} \label{defPM} {\bf (Perfect matching (PM))}
Let $i_1,\ldots,i_n\in [n]$ be row indices that are all different, i.e.,
$\{i_1,\ldots,i_n\} \= \{1,\ldots,n\}$.
Then the vector 
\begin{equation}\label{M1}
M\= \lor(i_1,1), (i_2,2),\ldots,(i_n,n)\ler
\end{equation}
is called a {\em perfect matching (PM)}. For short notation, we collect the indices $i_1,\ldots,i_n$ 
in a vector $\underline i \= (i_1,\ldots,i_n)$, and refer to $M$ as the PM generated by $\underline i$. This relationship is expressed by the notation
$M=M(\underline i)$.
\end{definition}
Thus, a perfect matching in an $n\times n$ matrix selects the indices of $n$ matrix entries, such that they all fall in 
different rows and different columns. We list them in a column-by-column order, but this choice is only made for simplicity, they could be listed in 
any order. 

Let $D$ denote the subset of $[n]^n$ consisting of vectors with all different coordinates. That is,
$$
D\= \{\underline i\;|\; \underline i =(i_1,\ldots,i_n)\in [n]^n,\; \forall k,j\in [n]:\; k\neq j \, \Rightarrow \, i_k\neq i_j\}.
$$
As a direct consequence of Definition~\ref{defPM}, with the introduced  notation we have that $\underline i \in D$ if and only if 
$M({\underline i})$ is a PM in $X$.

\begin{definition} \label{defPMVV} {\bf (Perfect matching value vector (PMVV))}
If $ M\= \lor(i_1,1), (i_2,2),\ldots,(i_n,n)\ler $
is a perfect matching, then the vector $$X_M \= (X_{i_1,1}\ldots,X_{i_n,n})$$ is called 
the corresponding {\em perfect matching value vector (PMVV)}.
\end{definition}
Observe that if $M$ is a PM, then it uniquely determines a PMVV $X_M$. On the other hand, the mapping between PMs and PMVVs is not necessarily 1--1:
depending on the values of matrix entries, it may happen that $M\neq M'$, yet $X_M=X_{M'}$.

\begin{definition} \label{defHPM} {\bf ($H$-perfect matching ($H$-PM))}
Let $H\subseteq S$ be a non-empty subset of $S$.  If a $M$ is a PM, and each component of the corresponding value vector $X_M$ falls in $H$, then 
$M$ is called an {\em $H$-perfect matching ($H$-PM)}. In this case the corresponding value vector is referred to as an {\em $H$-perfect matching value vector 
($H$-PMVV).}
\end{definition}

\noindent
For short reference, we call a matrix entry that falls in $H$  an {\em $H$-entry}. With this terminology,
an $H$-PM is a PM whose PMVV consists of $H$-entries only.

Note that if $H=S$, then an $H$-PM is simply the same as a PM.
Observe  that every square matrix has a PM, but not necessarily an $H$-PM. For example,
the main diagonal $\lor(1,1), (2,2),\ldots,(n,n)\ler $ always gives us a PM. On the other hand, when $H\neq S$, 
an $H$-PM may or may not exist.
In the special case, when $X$ is the bi-adjacency matrix of a balanced bipartite graph and $H=\{1\}$, then an $H$-PM corresponds to
a perfect matching in the bipartite graph in the usual graph theoretic sense. This motivates the name.

To shorten notations, we use the following convention. Let $A_1,\ldots,A_n$ be arbitrary sets. Define a {\em set vector} as a vector 
containing the sets as components: $A=(A_1,\ldots,A_n)$. For a vector  $y=(y_1,\ldots,y_n)$
the relationship that $y_i\in A_i$ holds 
for every $i\in [n]$ is denoted by  $y \;\underline \in\; A$. Note that this is just a shorter notation for 
$y\in A_1\times\ldots\times A_n$. Further, for the set $H$, we use $V_H$ to denote the following special set vector:
$$
V_H \= (\underbrace{H,\ldots,H}_{n\; {\rm times}})
$$ 
With the introduced  notations we can summarize  what makes a vector of the form (\ref{M1}) an $H$-PM:
\begin{quote}
For any index vector $\underline i =(i_1,\ldots,i_n)\in [n]^n$, the vector $M({\underline i})\= \lor(i_1,1), (i_2,2),\ldots,(i_n,n)\ler$
becomes an $H$-PM if and only if both $\underline i \in D$ and $X_{M(\underline i)} \;\underline\in\; V_H$ hold.
\end{quote}

If random row indices are used,  we denote them by $\nu_1,\ldots,\nu_n$. That is, 
in place of the index vector  $\underline i =(i_1,\ldots,i_n)$, the random index vector $\underline \nu =(\nu_1,\ldots,\nu_n)$ is used, 
and whenever $\underline \nu\in D$, it generates a random PM $M({\underline \nu})$. 
If $X_{M(\underline \nu)} \;\underline\in\; V_H$ also holds, then it is a random $H$-PM.
The  probability distribution of $X_{M(\underline \nu)}$ depends on the distribution of $\underline \nu$. 

Note that, in general, when we call an object random,
it does not necessarily mean that it is chosen {\em uniformly} at random from the set of corresponding objects. Whenever it is chosen uniformly at random,
we call it a uniform random object (for example, a uniform random $H$-PM in a matrix). When the domain from which the random object is chosen is also random,
then uniformity means that the choice is uniformly random for any fixed realization of the domain. For example, a uniform random $H$-PM in a random matrix
means that given any realization of the matrix, the $H$-PM is chosen uniformly at random from this realization. 


\subsection{The Independence Property}
\label{indprop}

Using the  concepts and notations introduced in Section~\ref{noti}, we now prove an important result.

\begin{theorem} \label{product} {\bf (Independence property of uniform random  $H$-perfect matching)}
Let $X$ be an $n\times n$-sized  random matrix, with entries from a common domain $S$. 
Let $H\subseteq S$ be a subset and $\alpha$ be a probability distribution on $H$, such that 
$X$ is $(H,\alpha)$-anchored and has an $H$-PM. Let $\cal M$ be a uniform random $H$-PM in $X$.
Then the components of the corresponding random value vector $X_{\cal M}$ are independent.
Furthermore, the distribution of $X_{\cal M}$  has the following product-form representation: for every fixed vector
$a=(a_1,\ldots,a_n) \;\underline\in\; V_H$  it holds that 
\begin{equation}\label{prodeq}
\Pr(X_{\cal M}=a) \= \prod_{k=1}^n \alpha(a_k).
\end{equation}
\end{theorem} 

\noindent
{\bf Remark.} The message of Theorem~\ref{product} is the following (surprising) statement: If $X$ has an $H$-PM, then the 
joint distribution of the components in the corresponding value vector of a uniform random $H$-PM is the same as what one would obtain by 
drawing each component  
 {\em independently} at random from the set $H$, each from the distribution $\alpha$. 
 Note that the component  {\em positions} in the matrix cannot be independent of each other,
 since in a matching they are forced to fall into different rows and columns. Further, whether certain entries constitute 
 an $H$-perfect matching depends on the random matrix, so ${\cal M}$ and $X$ are generally not independent, either. 
But the {\em values} in $X_{\cal M}$ are still independent of each other, assuming that the underlying random matrix is  $(H,\alpha)$-anchored
(see Definition~\ref{conform}). At the same time,
the $(H,\alpha)$-anchored property does not imply the independence of all matrix entries, so the independence of the $H$-PMVV components may 
indeed sound surprising.

To prove Theorem~\ref{product}, we first need a simple lemma, which claims that a uniform random variable, if  conditioned on falling into any (nonempty) subset of its range, remains uniform on the subset. For our purposes it is enough to consider finite ranges, so for simplicity we stay with that.

\begin{lemma}\label{uniform}
Let $\xi$ be a random variable taking its values from a finite set $S$, and let $S_1\subseteq S$ be a nonempty subset.
Let $\eta$ be another random variable with range $S_1$. We say that $\eta$ is the restriction of $\xi$ to $S_1$ if for any $x\in S_1$ it holds that 
$\Pr(\eta=x) \= \Pr(\xi=x\;|\; \xi\in S_1)$. Under these conditions, if $\xi$ has uniform distribution over $S$, then $\eta$ also has 
uniform distribution over $S_1$.
\end{lemma}

\noindent
{\bf Proof.} It is clear that $\eta$ is indeed  a random variable with range $S_1$, since $\Pr(\eta=x) \= \Pr(\xi=x\;|\; \xi\in S_1) \geq 0$ 
holds for every $x\in S_1$, and if $x\in S_1$ then 
$$
\sum_{x\in S_1} \Pr(\eta=x) \= \sum_{x\in S_1} \Pr(\xi=x\;|\; \xi\in S_1) \= \sum_{x\in S_1} \frac{ \Pr(\xi=x, \xi\in S_1)}{\Pr(\xi\in S_1)} \=
$$
$$ 
\sum_{x\in S_1} \frac{ \Pr(\xi=x)}{\Pr(\xi\in S_1)} \= \frac{1}{\Pr(\xi\in S_1)}\sum_{x\in S_1} \Pr(\xi=x) \= \frac{1}{\Pr(\xi\in S_1)}\Pr(\xi\in S_1) \= 1.
$$
Here in the third equality we used that $\xi=x$ implies $\xi\in S_1$, due to $x\in S_1$, therefore, the equality $\Pr(\xi=x, \xi\in S_1)=\Pr(\xi=x)$ holds. 
We also used that $S_1\neq \emptyset$, implying $\Pr(\xi\in S_1)>0$, due to the uniform distribution of $\xi$, thus making the denominators strictly positive.

Now let $x,y\in S_1$. Then we can write
\begin{equation}\label{etax}
\Pr(\eta=x) \= \Pr(\xi=x\;|\; \xi\in S_1) \= \frac{\Pr(\xi=x, \xi\in S_1)}{\Pr(\xi\in S_1)} \= \frac{ \Pr(\xi=x)}{\Pr(\xi\in S_1)}.
\end{equation}
Similarly, for $y\in S_1$ we have
\begin{equation}\label{etay}
\Pr(\eta=y) \= \Pr(\xi=y\;|\; \xi\in S_1) \= \frac{\Pr(\xi=y, \xi\in S_1)}{\Pr(\xi\in S_1)} \= \frac{ \Pr(\xi=y)}{\Pr(\xi\in S_1)}.
\end{equation}
Observing now that $\Pr(\xi=x)=\Pr(\xi=y)$ holds by the assumed uniform distribution of $\xi$, we get from (\ref{etax}) and (\ref{etay}) that
$\Pr(\eta=x)=\Pr(\eta=y)$ also holds for any $x,y\in S_1$. Thus, $\eta$ has uniform distribution over $S_1$, as claimed.

\hspace*{10mm} \hfill $\spadesuit$

\noindent
{\bf Remark.} While Lemma~\ref{uniform} expresses a simple property, it may lead to nontrivial facts when applied to a complicated and 
implicitly defined subset $S_1$, as we are going to see in the proof of Theorem~\ref{product}.

\medskip\medskip

\noindent
{\bf Proof of Theorem~\ref{product}.}
 Beyond the notations introduced in Section~\ref{noti}, a further short notation we use is $\alpha(A_i)=\sum_{x\in A_i} \alpha(x)$. 
We are going to prove
that for any set vector $A=(A_1,\ldots,A_n)$ where $A_i\subseteq H$, $i\in [n]$, the following holds:
if $\cal M$ is a uniform random $H$-PM in $X$, then
\begin{equation}\label{prod1}
\Pr(X_{\cal M} \uin A) \= \prod_{k=1}^n \alpha(A_k).
\end{equation}
This will imply both the claimed independence and the product-form distribution (\ref{prodeq}).

To prove (\ref{prod1}), recall first from Definition~\ref{defPM} that for an index vector $\underline \nu = (\nu_1,\ldots,\nu_n)\in D$ (random or not)
we have the perfect matching
$$M(\underline \nu)\=\lor(\nu_1,1),\ldots,(\nu_n,n)\ler. $$ 
Further, the definition directly implies that $\underline \nu$ and $M(\underline \nu)$ determine each other in a 1--1 way.
Therefore, $M(\underline \nu)$ becomes a uniform random PM if and only if $\underline \nu$ is a uniform random vector over $D$.

Let us use the above observation to derive a different, but equivalent formulation of $\Pr(X_{\cal M} \uin A)$, where $\cal M$ is a 
uniform random $H$-PM in $X$.
By  Definition~\ref{defHPM}, if $M$ is {\em any} PM in $X$, then it becomes an $H$-PM if and only if $X_M\uin V_H$ also holds. Observe 
further that under the condition $X_M\uin V_H$, a uniform random PM in $X$ also becomes a uniform random $H$-PM in $X$. The reason is that 
the condition $X_M\uin V_H$ defines a subset of the  range of $M$, i.e., a subset of all PMs in $X$. 
We know that  $M$ is selected uniformly at random from among all PMs, and we also know from Lemma~\ref{uniform} that whenever a uniform random variable is 
conditioned on falling into any (nonempty) subset of its range, 
then it remains uniform on this subset. 
The non-emptiness of the subset follows from the assumption that $X$ has an $H$-PM. Therefore, if 
$\underline \nu$ is a uniform random vector over $D$, then the corresponding  $M(\underline \nu)$, conditioned on
$X_{M(\underline\nu)} \uin V_H$, becomes a uniform random $H$-PM in $X$. Thus, we can reformulate the left-hand side of (\ref{prod1}) as follows.
Let $\underline \nu$ be a uniform random index vector over $D$, independent of $X$ (we are going to use this independence later). 
Then we can write: if $\cal M$ is a uniform random $H$-PM in $X$ and $\underline \nu$ is a uniform random index vector over $D$, independent of $X$, then
it holds that 
\begin{equation}\label{X_M}
\Pr(X_{\cal M} \uin A) \= \Pr \lor X_{M(\underline\nu)} \uin A \;|\; X_{M(\underline\nu)} \uin V_H \ler,
\end{equation}
where $A$ is the set vector used in (\ref{prod1}).

We are going to proceed by 
computing the right-hand side of (\ref{X_M}). For this purpose, let us use the law of total probability with the 
disjoint and exhaustive system of events $\{\underline\nu=\underline i\},\, \underline i\in D$. This system of events is indeed 
disjoint, as $\underline \nu\in D$ can take precisely one $\underline i$ value at a time, and it is also 
exhaustive, as $\underline i$ runs over $D$, which is the range of $\underline\nu$.
Thus, we get for the right-hand side of (\ref{X_M}):
\begin{eqnarray}\label{prod2}
 \Pr \lor X_{M(\underline\nu)} \uin A \;|\; X_{M(\underline\nu)} \uin V_H \ler &=&
\sum_{\underline i\in D}  \Pr \lor X_{M(\underline\nu)} \uin A \;|\; X_{M(\underline\nu)} \uin V_H, \; \underline \nu = \underline i \ler
\times \\ \nonumber
& & \Pr\lor \underline \nu=\underline i\;|\; X_{M(\underline\nu)} \uin V_H \ler 
\end{eqnarray}
Let us now take a closer look at the probability 
$\Pr \lor X_{M(\underline\nu)} \uin A \;|\; X_{M(\underline\nu)} \uin V_H,  \underline \nu = \underline i \ler$
on the right-hand side of (\ref{prod2}). Due to the condition $\underline \nu = \underline i$, all other occurrences of $\underline \nu$
in this probability can be replaced by $\underline i$, giving us 
\begin{eqnarray}\label{prod3}
\Pr \lor X_{M(\underline\nu)} \uin A \;|\; X_{M(\underline\nu)} \uin V_H, \; \underline \nu = \underline i \ler \=
\Pr \lor X_{M(\underline i)} \uin A \;|\; X_{M(\underline i)} \uin V_H, \; \underline \nu = \underline i \ler.
\end{eqnarray}
Looking at the right-hand side of (\ref{prod3}), recall that $\underline \nu$ is a uniform  random vector over $D$, chosen {\em independently} of $X$. 
Therefore, for any fixed $\underline i\in D$ the vectors $X_{M(\underline i)}$ and $\underline \nu$ become independent. The reason is that 
$M(\underline i)$ is a fixed PM for a fixed $\underline i\in D$, so the value vector $X_{M(\underline i)}$ becomes a 
fixed sub-list of the matrix entries (see Definition~\ref{defPMVV}), which makes 
$X_{M(\underline i)}$ a function of $X$. As $X$ and $\underline \nu$ are assumed independent, and functions of independent random variables 
remain independent, therefore,  $X_{M(\underline i)}$ and $\underline \nu$ are indeed independent.

Using the independence of $X_{M(\underline i)}$ and $\underline \nu$, 
we can eliminate  $\underline \nu$ from the condition on the right-hand side of (\ref{prod3}):
\begin{eqnarray} \nonumber
\Pr \lor X_{M(\underline i)} \uin A \;|\; X_{M(\underline i)} \uin V_H,\,  \underline \nu = \underline i \ler &=&
\frac{ \Pr \lor X_{M(\underline i)} \uin A,\, X_{M(\underline i)} \uin V_H,\,  \underline \nu = \underline i \ler }
{\Pr \lor X_{M(\underline i)} \uin V_H,\,  \underline \nu = \underline i \ler} \\ 
&=& \frac{ \Pr \lor X_{M(\underline i)} \uin A,\, X_{M(\underline i)} \uin V_H\ler \Pr\lor \underline \nu = \underline i\ler }
{\Pr \lor X_{M(\underline i)} \uin V_H\ler  \Pr\lor \underline \nu = \underline i \ler} \\ \nonumber
&=& \frac{ \Pr \lor X_{M(\underline i)} \uin A,\, X_{M(\underline i)} \uin V_H\ler }
{\Pr \lor X_{M(\underline i)} \uin V_H\ler } \\ \nonumber
&=& \Pr \lor X_{M(\underline i)} \uin A \;|\; X_{M(\underline i)} \uin V_H \ler.
\end{eqnarray}
This allows the  simplification of   the right-hand side of (\ref{prod3}) to 
\begin{equation}\label{prod4}
\Pr \lor X_{M(\underline i)} \uin A \;|\; X_{M(\underline i)} \uin V_H, \; \underline \nu = \underline i \ler \=
\Pr \lor X_{M(\underline i)} \uin A \;|\; X_{M(\underline i)} \uin V_H \ler.
\end{equation}
To make sure that all these expressions are well defined, we also need to show that the following hold for every $\underline i\in D$: 
\begin{description}

\item[\,\,\,\,\,\,\,\, \rm (i)]  $\, \Pr(\underline \nu = \underline i)>0$

\item[\,\,\,\,\,\,\, \rm (ii)] $\Pr(X_{M(\underline i)} \uin V_H)>0$ 

\item[\,\,\,\,\,\, \rm (iii)] $\Pr(X_{M(\underline i)} \uin V_H, \; \underline \nu = \underline i)>0$. 

\end{description}
Here (i) follows from the facts that $\underline i \in D$, and $\underline \nu $ is uniform over $D$,
yielding  $\Pr(\underline \nu = \underline i)=\frac{1}{|D|}>0$.
The claim (ii)  can be proved by writing for any $\underline i\in D$
\begin{equation}\label{condi}
\Pr(X_{M(\underline i)} \uin V_H) \= \Pr(X_{i_1,1}\in H,\ldots, X_{i_n,n}\in H) \= \prod_{k=1}^n \Pr(X_{i_k,k}\in H) \;>\; 0,
\end{equation}
where the product form follows from the independence of $X_{i_1,1},\ldots,X_{i_n,n}$, as they fall in different rows, and the rows are independent, 
due to the assumed $(H,\alpha)$-anchored property of $X$ (see also the note below). The positive value of the product is implied by the assumption that 
$\Pr(X_{ij} \in H)>0$ holds for every $i, j\in [n]$, which is also part of the $(H,\alpha)$-anchored property. Finally, (iii) is a consequence 
of (i) and (ii), as well as the fact that for any fixed $\underline i$
the vectors $X_{M(\underline i)}$ and $\underline \nu$ are independent, since $\underline \nu\in D$ was chosen independently of the random matrix.

\begin{quote}
{\em Note:} If in the proof of (ii) we want to more formally derive the independence of $X_{i_1,1},\ldots,X_{i_n,n}$ from the $(H,\alpha)$-anchored property, 
then we can argue as follows.
Let us use the well known fact that functions of independent random variables remain independent. Let $f_k$ be 
the function that extracts the $k^{\rm th}$ coordinate from any vector, that is, if $v=(v_1,\ldots,v_n)$, then $f_k(v)=v_k$. Then, denoting the 
row vectors of $X$ by $X^{(1)},\ldots,X^{(n)}$, we get
\begin{eqnarray}\nonumber
X_{i_1,1} &=& f_1\lor X^{(i_1)}\ler \\ \nonumber
X_{i_1,2} &=& f_2\lor X^{(i_2)}\ler  \\ \nonumber
\vdots \\ 
X_{i_1,n} &=& f_n\lor X^{(i_n)}\ler \label{functions}
\end{eqnarray}
Since $i_1,\ldots,i_n$ are all different, due to $\underline i\in D$, and they are constant indices, 
therefore, the 
row vectors $X^{(i_1)},\ldots,X^{(i_n)}$ constitute distinct rows in fixed positions, which, by the $(H,\alpha)$-anchored property, makes them independent. 
Then their functions $X_{i_1,1},\ldots,X_{i_n,n}$, as given in (\ref{functions}),  are also independent.
\end{quote}

Let us now further analyze the conditional probability $\Pr(X_{M(\underline i)}\uin A\;|\; \, X_{M(\underline i)} \uin V_H)$. 
We can write
\begin{eqnarray}\nonumber
\Pr(X_{M(\underline i)}\uin A\;|\; X_{M(\underline i)} \uin V_H)
&=& \Pr(X_{i_1,1}\in A_1,\ldots,X_{i_n,n}\in A_n\;|\; X_{i_1,1}\in H,\ldots,X_{i_n,n}\in H) \\[2mm] 
&=& \frac{ \Pr(X_{i_1,1}\in A_1,\ldots,X_{i_n,n}\in A_n, X_{i_1,1}\in H,\ldots,X_{i_n,n}\in H)} 
         {\Pr(X_{i_1,1}\in H,\ldots,X_{i_n,n}\in H)}.  \nonumber
\end{eqnarray}
Using that $\forall k: A_k\subseteq H$, we get  $\Pr(X_{i_k,k}\in A_k,X_{i_k,k}\in H) =\Pr(X_{i_k,k}\in A_k)$ for any $k$. Therefore, we can simplify 
the above expression of $\Pr(X_{M(\underline i)}\uin A\;|\; X_{M(\underline i)} \uin V_H)$ as
\begin{equation}\label{long2}
\Pr(X_{M(\underline i)}\uin A\;|\; \, X_{M(\underline i)} \uin V_H) \=
\frac{ \Pr(X_{i_1,1}\in A_1,\ldots,X_{i_n,n}\in A_n)} 
         {\Pr(X_{i_1,1}\in H,\ldots,X_{i_n,n}\in H)}.
\end{equation}
Consequently, due to the already known independence of  $X_{i_1,1},\ldots,X_{i_n,n}$,
the probabilities on the right-hand side of (\ref{long2}) can be written in product form, resulting in
\begin{eqnarray}\label{long3} \nonumber
\Pr(X_{M(\underline i)}\uin A\;|\; \, X_{M(\underline i)} \uin V_H)
&=& \frac{ \Pr(X_{i_1,1}\in A_1)\times\ldots\times \Pr(X_{i_n,n}\in A_n)} 
         {\Pr(X_{i_1,1}\in H)\times\ldots\times \Pr(X_{i_n,n}\in H)} \\[2mm]
&=& \frac{ \Pr(X_{i_1,1}\in A_1)}{\Pr(X_{i_1,1}\in H)} \times\ldots\times \frac{\Pr(X_{i_n,n}\in A_n)} {\Pr(X_{i_n,n}\in H)}. 
\end{eqnarray}
Observe that by the $(H,\alpha)$-anchored property of $X$, the fractions in the last row of (\ref{long3}) satisfy
$$
\forall k\in [n]: \;\;\;\;\;\frac{ \Pr(X_{i_k,k}\in A_k)}{\Pr(X_{i_k,k}\in H)} \= \alpha(A_k).
$$
Substituting  this into  (\ref{long3}), we get
\begin{equation}\label{prod5}
\Pr(X_{M(\underline i)}\uin A\;|\; \, X_{M(\underline i)} \uin V_H) \= \prod_{k=1}^n \alpha(A_k).
\end{equation}
Recall now that (\ref{prod3}) and (\ref{prod4}) together give 
$$
\Pr \lor X_{M(\underline\nu)} \uin A \;|\; X_{M(\underline\nu)} \uin V_H,  \underline \nu = \underline i \ler \=
\Pr \lor X_{M(\underline i)} \uin A \;|\; X_{M(\underline i)} \uin V_H \ler.
$$
With (\ref{prod5}), this yields
$$
\Pr \lor X_{M(\underline\nu)} \uin A \;|\; X_{M(\underline\nu)} \uin V_H,  \underline \nu = \underline i \ler \=
\prod_{k=1}^n \alpha(A_k).
$$
Substituting the resulting expression of  
$\Pr \lor X_{M(\underline\nu)} \uin A \;|\; X_{M(\underline\nu)} \uin V_H,  \underline \nu = \underline i \ler$ 
into the right-hand side of (\ref{prod2}), we obtain
\begin{equation}\label{prod6}
 \Pr \lor X_{M(\underline\nu)} \uin A \;|\; X_{M(\underline\nu)} \uin V_H \ler \=
  \prod_{k=1}^n \alpha(A_k) \underbrace{\sum_{\underline i\in D}
\Pr\lor \underline \nu=\underline i\;|\; X_{M(\underline\nu)} \uin V_H \ler}_{=1} \= \prod_{k=1}^n \alpha(A_k).
\end{equation}
Note that the conditional probabilities are well defined, as $\Pr(X_{M(\underline\nu)} \uin V_H)>0$ holds, due to  (\ref{condi}),
which applies for every $\underline i\in D$. Now, taking (\ref{X_M}) into account, we obtain that if $\cal M$ is a uniform random $H$-PM, 
and $A=(A_1,\ldots,A_n)$ with $A_i\subseteq H$, $i\in [n]$, then 
\begin{equation}\label{prod7}
\Pr(X_{\cal M} \uin A) \= \prod_{k=1}^n \alpha(A_k)
\end{equation}
holds, which was set as a goal in (\ref{prod1}) at the beginning of the proof. 

This directly implies the claimed product-form representation (\ref{prodeq}) by choosing $A=(A_1,\ldots,A_n)$ with
$A_1=\{a_1\},\ldots,A_n=\{a_n\}$. To show the independence 
of the components of $X_M$, we need slightly more. At the first sight, the product-form distribution (\ref{prodeq}) appears equivalent to independence, 
but in fact it is weaker, because  independence requires  that such a product-form distribution should  also hold 
for every {\em subset} of the random variables, not only for the whole set\footnote{Counterexamples are known when the product-form 
distribution holds for the whole set of certain random variables, 
but not for all subsets, and they are indeed not independent (see, e.g., Stoyanov \cite{stoyanov}).}. This, however, also follows from 
(\ref{prod7}). If we want a product-form representation for a subset $\{i_1,\ldots,i_r\}$ of indices, then 
we can take $A_{i_j}=\{a_{i_j}\}, \, j=1,\ldots,r,$ and $A_k=H$ whenever $k\notin \{i_1,\ldots,i_r\}$. It will eliminate the
undesired indices from the product by setting the corresponding factors to unity via 
$\Pr\lor X_{\nu_k,k} \in A_k\;|\;X_{\nu_k,k} \in H \ler = \alpha(A_k) = \alpha(H)=1$, which indeed holds with $A_k=H$,
taking into account  that the domain of $\alpha$ is $H$.

\hspace*{10mm} \hfill $\spadesuit$

\section{The Almost Uniform Sampler}
\label{almost}

Our approach is based on creating a probabilistic polynomial time algorithm for the almost uniform {\em sampling} of independent sets in any graph. 
Then, by known results, it will imply the existence of an FPRAS for the approximate 
{\em counting} of independent sets in bounded degree graphs, which in turn has very surprising consequences, including ${\bf  NP}={\bf  RP}$.

Below we provide a description of a fully polynomial almost uniform sampler of independent sets in arbitrary graphs. 
Let us first define what it is that we are looking for.

\begin{definition} \label{eaus}
{\bf (Fully polynomial almost uniform sampler of independent sets.)} A probabilistic algorithm $\cal A$ is called a 
 {\em fully polynomial almost uniform sampler} of independent sets in a graph, if the following hold:
\vspace*{-2mm}
\begin{itemize}

\item 
The input of $\cal A$ is an undirected simple graph $G$ on $N$ vertices and $m$ edges. 
Further, a rational error parameter $\vareps\in (0,1]$ is also given as part of the input.

\item The output of $\cal A$ is a random independent set $A\in {\cal I}(G)$, such that its probability distribution 
over ${\cal I}(G)$, denoted by $\beta$, 
satisfies $d_{TV}(\beta, U) \leq \vareps$, where $d_{TV}$ is the total variation distance, and $U$ is the uniform distribution over ${\cal I}(G)$. 

\item The  running time of $\cal A$ is bounded by a polynomial in $N$ and $\log(1/\vareps)$.

\end{itemize}

\end{definition}

\noindent
Note: The adjective ``fully polynomial" refers here to the fact that the running time is polynomial in $\log(1/\vareps)$, so it is preserved even with 
an exponentially small error bound $\vareps$. 

Below we present an algorithm that satisfies the requirements of Definition~\ref{eaus}. We also include some comments in the 
description of the algorithm, to make it easier to follow. 

The algorithm includes three counter variables that are used to count certain quantities, 
as follows:

\begin{itemize}

\item {\em Repetition counter,} denoted by $r$. We repeat the algorithm several times (with independent randomness), the variable $r$ serves to identify 
the repetitions. Its maximum value is $R=\lceil \log_2 (2/\vareps) \rceil $.

\item {\em Round counter,} denoted by $k$. Within each repetition, we create successive approximations $G_0,G_1,\ldots,G_m$, of the input graph in Step 1.
In round $k$ we are going to deal with the graph $G_k$.

\item {\em Failure counter}, denoted by $F$. What we call failure is when the current random matrix $X$ does not have an $H$-PM. 
These occurrences are counted by $F$. Whenever failure happens in a repetition, we do not continue that repetition, so there can be at most one failure per
repetition. 

\end{itemize}

\medskip\medskip

\noindent
{\bf Algorithm 1: Fully polynomial almost uniform sampler of independent sets}



\begin{description}

\item[Step 1 (Creating successive approximations of the input graph)] 
Let $e_1,\ldots,e_m$ be the ed\-ges of $G$, in arbitrary order. 
Construct the graphs  $G_0,G_1,\ldots,G_m$, such that each graph has all the $N$ vertices of $G$, and 
for every $k=0,1,\ldots,m$ the graph $G_k$ contains the first $k$ edges of $G$. That is, $G_0$ is the empty graph, $G_1$ keeps only the edge $e_1$,
$G_2$ keeps edges $e_1,e_2$, and so on. The last graph $G_m$ contains all edges, so it becomes the same as the original input graph $G$.

Set $n=2N^2$ ($n$ is the number of Markov chains we are going to run in parallel in each repetition). Further,
set the maximum repetition count to $R=\lceil \log_2 (2/\vareps) \rceil $.

Set the initial values of the repetition counter to $r=1$, and of the failure counter to $F=0$.

\item[Step 2 (Round 0: Initialization for repetition $r$)]  
Pick $n$ uniformly random independent sets $A_1,\ldots,A_n\in {\cal I}(G_0)$, so that they are 
independent of each other as random variables. \\ 
{\em Comment:} in $G_0$ this 
is trivial to accomplish, as the graph consists of isolated vertices. That is, in $G_0$ we may simply put each vertex into $A_i$
independently with probability $\frac{1}{2}$, for each $i$.

Set the round counter to $k=1$.

\item[Step 3 (Running the Markov chain in round $k$)] Run $n$ realizations of the BIDC Markov chain independently in $G_k$ (for the description of BIDC see Section~\ref{BIDC}).
Start the $i^{th}$ realization from 
initial state $A_i, \; i=1,\ldots,n$. Run each realization for $n$ steps, and let $X_{i0}, X_{i1},\ldots,X_{in}$ denote the state 
sequence taken by the $i^{th}$ realization, where $X_{i0}=A_i$. 

\item[Step 4 (Creating the matrix)] Collect the $n$ realizations in an $n\times n$-sized matrix $X=[X_{ij}]$, 
so that the  $i^{th}$ realization is filled into the $i^{th}$ row, 
$ i\in [n]$, starting the row with $X_{i1}$. Set $S={\cal I}(G_k)$ as the domain containing all matrix entries,  and set $H={\cal I}(G_{k+1})\subseteq S$
as the target set.

\item[Step 5 (Failure check)] Check if there is an $H$-PM in $X$ with $H={\cal I}(G_{k+1})$. (For the definition of $H$-PM see Definition~\ref{defHPM}; for 
checking the existence of an $H$-PM, see Algorithm 2 in Section~\ref{HPM}.) 
\begin{itemize}

\item 
If there is an $H$-PM in $X$ (no failure), then continue from Step 6.

\item If there is no $H$-PM in $X$ (failure), then increment the failure counter $F$. 

\item If $F=r<R$ (i.e., all repetitions failed so far, but the maximum number of repetitions has not been reached yet), then
increment $r$, set $k=1$, and  start over from Step 2 (i.e., move to the next repetition). 

\item If $F=R$ (i.e., each repetition failed, and the maximum is reached), then set $A_1=\{v_0\}$,
where $v_0$ is a fixed vertex of $G$; output $A_1$ and {\sc halt}.

\end{itemize}

\item[Step 6 (Finding a random $H$-PM)] 
Find an approximately uniform random $H$-PM $\cal M$ in $X$, such that its distribution deviates from the uniform distribution
by at most $\delta$ in total variation distance, 
with $\delta=2^{-n-Rm-1}$, where $n=2N^2$, as well as $N,m$ are the number of vertices and edges, respectively, in the input graph $G$, 
and $R$ is the maximum repetition count. 
To find such an $\cal M$ in probabilistic  polynomial time, see Algorithm 3 in Section~\ref{HPM}.

\item[Step 7 (Creating initial values for the next round)] Let $X_{\cal M}$ be the $H$-PM value vector  corresponding to $\cal M$ 
(see Definitions \ref{defPMVV}, \ref{defHPM}).
For every $i=1,\ldots,n$, let the new value of $A_i$ be the $i^{th}$ component of $X_{\cal M}$.

\item[Step 8 (Loop, or output and halt)] If $k<m$ (i.e., rounds still remain in the current repetition), then set $k=k+1$, and go to Step 3. Else 
(if the current repetition is completed  without failure), then output $A_1$ and {\sc halt}.





\end{description}

\subsection{Algorithms Related to $H$-Perfect Matchings}
\label{HPM} 

In Step 5 of Algorithm 1 we need to check whether the matrix contains an $H$-perfect matching, and in Step 6 we look for an approximately uniform random 
$H$-perfect matching. In this section we detail how these tasks can be carried out.

Recall that an $H$-perfect matching ($H$-PM) in a matrix $X$ is a perfect matching (PM) in $X$ (see Definition~\ref{defPM}), with the additional
property that the involved matrix entries all fall in the subset $H$. In other words, the value vector of the PM (see Definition~\ref{defPMVV})
consists of $H$-entries only. To reduce the $H$-PM to the ordinary matching problem in bipartite graphs, we introduce the following definition:

\begin{definition}\label{H-skel} {\bf ($H$-skeleton of matrix)}
Let $X=[X_{ij}]$ be an $n\times n$-sized matrix, with entries from a common domain $S$. 
Let $H\subseteq S$ be a subset. The {\em $H$-skeleton} of $X$ is a $n\times n$-sized 0--1 matrix, denoted by $X^H=[X^H_{ij}]$, 
such that the entries of $X^H$ are defined by 
\begin{eqnarray} \nonumber
X^H_{ij} = \left\{
    \begin{array}{ccl}
      1  & \;\;\mbox{ if }\;\;\; X_{ij}\in H \\
      0  & \;\;\mbox{ if }\;\;\; X_{ij}\notin H.
    \end{array} \right.
\end{eqnarray} 
\end{definition}

Using the $H$-skeleton, we can easily connect the existence of an $H$-PM to the existence of an ordinary matching in a bipartite graph, as stated 
in the following lemma.

\begin{lemma} \label{lemma2} Let $X=[X_{ij}]$ be an $n\times n$-sized matrix, with entries from a common domain $S$, and  
let $H\subseteq S$ be a subset. Then $X$ has an $H$-PM if and only if the $2n$-vertex  balanced bipartite graph $G_B$ has a perfect matching 
(in the ordinary sense), where $B=X^H$ is the $H$-skeleton of $X$, as defined in {\em Definition~\ref{H-skel},} and $B$ serves as the bi-adjacency matrix
of $G_B$.
\end{lemma}

\noindent {\bf Proof.}
Set $H_1=\{1\}$. Then by Definition~\ref{defHPM}, the 0--1 matrix $B=X^H$ has an $H_1$-PM if and only if the bipartite graph $G_B$ has a perfect matching 
in the ordinary sense. Further, replacing each 1-entry in $B$ by the original $H$-entry of $X$ in the same position, we get that 
$B$ has an $H_1$-PM if and only $X$ has an $H$-PM. Thus, $X$ has an $H$-PM if and only if the balanced bipartite graph $G_B$ has a perfect matching 
in the ordinary sense.

\hspace*{10mm} \hfill $\spadesuit$

Deciding whether a bipartite graph has a perfect matching is a classical, well solved problem, for which a number of (deterministic) polynomial time algorithms are known. A well-known  solution is the Hopcroft-Karp algorithm, which constructs a maximum matching in a bipartite graph, 
see Hopcroft and Karp \cite{hopcroft}. It runs in $O(m\sqrt{n})$ time in a graph of 
$n$ vertices and $m$ edges, which is $O(n^{5/2})$ in the worst case, published in 1973. Later various improvements were found, including the  
$O(m^{10/7})$ algorithm of Madry \cite{madry} in 2013; and the $O(m^{4/3})$ algorithm of Liu and Sidford \cite{liu} in 2020. Eventually, an almost linear time algorithm was published quite recently (in 2022) by Chen, Kyng, Liu, Peng, Gutenberg, and Sachdeva \cite{chen}.

With a maximum matching algorithm, using Lemma~\ref{lemma2}, we can directly check if $X$ has an $H$-PM, to carry out Step 5 of Algorithm 1, as follows.

\smallskip
\noindent {\bf Algorithm 2: Deciding the existence of an $H$-PM in $X$} 
\vspace*{-2mm}
\begin{itemize} \itemsep=1mm

\item 
Create the $H$-skeleton $B=X^H$ of $X$ (see Definition~\ref{H-skel}).

\item Run  a maximum matching algorithm on the bipartite graph $G_B$ to obtain a maximum matching in  $G_B$.

\item If the obtained maximum matching is a perfect matching in $G_B$, then there is an $H$-PM in $X$,
otherwise there is none.

\end{itemize}

The other (harder) task in connection with $H$-PM is the one needed in Step 6 of Algorithm~1: finding an approximately uniform random $H$-PM in $X$.
Fortunately, we can use an existing result of Jerrum, Sinclair and Vigoda \cite{jerrum}, which they originally developed to approximate 
the permanent of a matrix with non-negative entries.
Their algorithm generates an approximately 
uniform random perfect matching in a bipartite graph. The approximate uniformity is measured here by the total variation distance
from the ideal uniform distribution.   Recall that the well known {\em total variation 
distance} between distributions $\alpha, \beta$ over a finite set $\Omega$ is 
$$
d_{TV}(\alpha, \beta) \= \sum_{x\in \Omega} |\alpha(x)-\beta(x)| \= \max_{A\subseteq \Omega} |\alpha(A)-\beta(A)|.
$$
The result of Jerrum, Sinclair and Vigoda (see \cite{jerrum}, Section 2) can be summarized in the following theorem:

\begin{theorem}\label{JSV} {\em (Jerrum, Sinclair and Vigoda \cite{jerrum})}
Let $G$ be a $2n$-vertex balanced bipartite graph that has a perfect matching, and $\delta>0$ be an error parameter. 
Let $\cal U$ denote the uniform distribution 
over the set of all perfect matchings in $G$. 
Then there exists a probabilistic polynomial time algorithm that outputs 
a random perfect matching $\cal M$  in $G$, such that $\cal M$ has distribution ${\cal D}$, which  satisfies 
$d_{TV}({\cal D}, {\cal U}) \leq \delta$, where $d_{TV}$ denotes the total variation distance. The running time of the algorithm is 
polynomial in $n$ and in $\log \delta^{-1}$.
\end{theorem}
{\em Note:} The actual running time of the algorithm  is $\widetilde O(n^{10}\log \delta^{-1})$, where $\widetilde O$ means that logarithmic factors of $n$ are 
suppressed. This can be improved to $\widetilde O(n^7\log \delta^{-1})$, see Bezakova, 
Stefankovic, Vazirani, and Vigoda \cite {bezakova}. For our main goal, however, 
at this point it only matters that the problem is solvable in time polynomial in  $n$ and in $\log \delta^{-1}$. 
This will be enough to prove ${\bf NP}={\bf  RP}$, we do not attempt to optimize the actual running time in the present paper.

Using Theorem~\ref{JSV}, we can produce an approximately uniform random $H$-PM in $X$ for Step 6 of Algorithm 1, as follows:

\smallskip
\noindent {\bf Algorithm 3:  producing an approximately uniform random $H$-PM in $X$} 
\vspace*{-2mm}
\begin{itemize} \itemsep=1mm

\item Create the $H$-skeleton $B=X^H$ of $X$ (see Definition~\ref{H-skel}).

\item Run the algorithm of Theorem~\ref{JSV} on the bipartite graph $G_B$ with error parameter $\delta=2^{-n-Rm-1}$,
where $n=2N^2$, as well as $N,m$ are the number of vertices and edges, respectively, of the original input graph $G$, and $R$ is the maximum repetition count.

\item  Let $\cal M'$ be the obtained approximately uniform random perfect matching in the bipartite graph $G_B$, and let 
${\cal M} = \lor(\nu_1,1),\ldots,(\nu_n,n)\ler$ denote the corresponding approximately uniform random $H_1$-PM in the 
bi-adjacency matrix $B$, with $H_1=\{1\}$.

\item Due to the $H$-skeleton construction, $\cal M$ is also an approximately uniform random $H$-PM in the matrix $X$. Return $\cal M$ to 
Step 6 of Algorithm 1.

\end{itemize}

\section{Analysis of Algorithm 1}
\label{analysis}

\begin{theorem}\label{main}  
Algorithm  1 implements a fully polynomial almost uniform sampler of independent sets, as defined in Definition~\ref{eaus}. 
\end{theorem}

\noindent
{\bf Proof of Theorem~\ref{main}.}
Consider repetition $r$, with $1\leq r\leq R$, and round $k$ within this repetition, $0\leq k\leq m$.
Let  $A^k=(A^k_1,\ldots,A^k_n)$ denote the initial states that are used in Step 3 to run $n$ realizations of the Markov chain in round $k$. 
When Step 3 is first executed, we have $k=1$.
At the end, when Step 3 is run for the last time,  $k=m$ holds (unless a failure occurred in Step 5).
In the first execution of  Step 3 with $k=1$, the initial sets $A^{1}=(A^{1}_1,\ldots,A^{1}_n)$ for $G_1$
can be determined in a trivial way, as noted in Step 2, since  the graph $G_0$ still 
consists of isolated vertices. Therefore,  to produce the independent, uniform system $A^{1}_1,\ldots,A^{1}_n$ of initial states for $G_1$,
we may simply put each vertex into $A^{1}_i$ independently with probability 1/2, for every $i$. 

For a general $k$, let us call the initial system $A^k=(A^k_1,\ldots,A^k_n)$ of round $k$ {\em ideal,} if each $A^k_i$ is perfectly uniform over 
${\cal I}(G_k)$, and they are independent of each other. For notational purposes, we distinguish such ideal systems with an asterisk in the lower index: 
if $A^k$ is ideal, then we denote it by $A_*^{k}=(A^{k}_{*1},\ldots,A^{k}_{*n})$.

First we analyze what happens if we use an ideal system $A_*^{k}$ in Step 3. 
Let $X$ be the matrix created in Step 4.
Running the BIDC Markov chain from the ideal initial states results in a matrix $X$ in which each entry is uniformly distributed over $S$, 
as the chain starts from its stationary distribution, so it must remain there. Further, the rows of $X$ are independent of each other, 
being created via independent realizations. The distribution of each entry when conditioned on falling in the subset 
$H={\cal I}(G_{k+1})\subseteq S$ remains uniform on $H$. The reason is that  it is uniform on $S$, so by Lemma~\ref{uniform} (see in Section~\ref{indprop}), 
we know that when the entry is conditioned on falling in any 
nonempty subset of $S$, it remains uniform on the subset\footnote{This conditioning is not to be confused with the first hit to $H$, 
which generally would not remain uniform on $H$.}.
Thus, if we choose $\alpha$ as the uniform distribution on $H$, then the matrix that results from the ideal input 
$A_*^{k}$ satisfies Definition~\ref{conform} (see Section~\ref{noti}), making the matrix  $(H,\alpha)$-anchored,  
with $H={\cal I}(G_{k+1})$.
This leads to the following:

\begin{quote}
{\bf Observation 1.}
The above properties of the obtained matrix $X$ allow us to apply the Independence Property (Theorem~\ref{product} in Section~\ref{indprop}), 
{\em if} two further conditions are also satisfied: (a) the matrix has an $H$-PM at all; and (b) given that there is an $H$-PM, we can generate a perfectly
uniform one. Once a perfectly uniform $H$-PM is available, then by the Independence Property (Theorem~\ref{product}) it would give us 
$n$ independent, perfectly uniform subset samples from $H$. 
This would allow us to start Step 3 in the  next round again from an ideal input $A_*^{k+1}$, by taking $A_*^{k+1}=X_{\cal M}$, where 
$X_{\cal M}$ is the value vector (see Definition~\ref{defPMVV}) of the perfectly uniform random $H$-PM in $X$.
\end{quote}
 
Let us take a closer look at how to handle  the two conditions (a) and (b) in Observation 1.
\begin{itemize}

\item 
Condition (a): Taking into account  that the different repetitions are independent of each other, it does not matter which repetition is considered,
as long as it is failure-free. Therefore, failures will cause a problem only if {\em each} repetition contains a failure.
We are going to prove that this bad case has exponentially small probability, 
introducing only a negligible deviation.

\item 
Condition (b): Given that there is an $H$-PM in $X$, 
we can obtain a random one from Algorithm 3. However, it is  is only {\em approximately} uniform, so we have to account 
for the deviation from the ideal case.

\end{itemize}

To handle these conditions, we analyze Algorithm 1 considering two cases:

{\em  (1) Well-behaved case.} In this case we assume that in at least one repetition an $H$-PM exists in $X$ in every round.

{\em (2) Deviant case.} In this case the Algorithm suffers failure in some round in {\em every} repetition, 
so it eventually outputs $\{v_0\}$ and halts in Step 5.

Let us analyze first the well-behaved case, in a repetition $r$ where no failure occurs.

Let $\cal M$ be the random $H$-PM in $X$, found in Step 6 of Algorithm 1.
This $H$-PM  $\cal M$ is produced by Algorithm 3 (see in Section~\ref{HPM}). The matching $\cal M$ has approximately uniform 
distribution over the set of all $H$-PMs of $X$.
The meaning of approximately uniform here is 
that the total variation distance between the distribution of $\cal M$ and the uniform distribution is at most $\delta$, 
where the parameter $\delta$ is set to $\delta=2^{-n-Rm-1}$. Here $n=2N^2$, where $N$ is the number of vertices in the input graph. 
Note that $\delta^{-1}$, which plays a role in the running time of Algorithm 3, depends exponentially 
on $n+Rm$. But this will not cause a problem, as the running time is polynomial in  $\log \delta^{-1}$, rather than in $\delta^{-1}$ (see Theorem~\ref{JSV}),
and $\log \delta^{-1}$ grows only  linearly with $n+Rm$.

Let $\cal D$ denote the distribution of $\cal M$ over the set of all $H$-PMs in $X$, and let $\cal U$ be the uniform distribution over the same set.
Observe that three types of perfect matchings are all in 1--1 correspondence here: 
(1) the $H_1$-PMs in the $H$-skeleton matrix $B=X^{H}$ with $H_1=\{1\}$ (see Definition~\ref{H-skel}, 
as well as Lemma~\ref{lemma2} with its proof);  (2) the perfect matchings in the corresponding 
bipartite graph $G_{B}$; and (3) the $H$-PMs in the matrix $X$. 
Therefore (with a slight abuse of notation), we can denote each distribution 
by $\cal D$, and use $\cal U$ for the ideal uniform distribution in all three cases.

We obtain from Theorem~\ref{JSV} that $\cal D,U$ are close to each other in terms of total variation distance: 
$d_{TV}({\cal D}, {\cal U}) \leq \delta$ holds. We also know from statement (vi) of Lemma~\ref{lemma1} that this yields
$s({\cal D}, {\cal U})\leq 2|S|\delta$ for the separation distance. Recall that $S$ is the family of independent sets in an $N$-vertex graph, implying
$|S|\leq 2^N$. Set $\rho=2^{n+1}\delta$. Then, due to our choice of $\delta=2^{-n-Rm-1}$, we get $\rho=2^{-Rm}$.
Putting these together,  we obtain the following upper bound:
\begin{equation}\label{sdu}
s({\cal D}, {\cal U})\leqq 2|S|\delta \leqq 2\cdot 2^N \delta \leqq 2^{n+1} \underbrace{2^{-n-Rm-1}}_{\delta} \= 2^{-Rm} \= \rho.
\end{equation}
Applying the Mixture Representation Property of the separation distance (see (iv) in Lemma~\ref{lemma1}), the above bound yields
\begin{equation}\label{mixD}
{\cal D} \= \left(1-\rho\right)\,{\cal U}\, + \, \rho \psi
\end{equation}
with some (unknown) distribution $\psi$. Consequently, the generation of $\cal M$ can be viewed as the following random event:
with probability $1-\rho$ we get a perfectly uniform random $H$-PM ${\cal M}_*$, and
with probability $\rho$ we obtain a random $H$-PM ${\cal M}_\psi$ from the unknown distribution $\psi$. Let us look into these two cases:
\begin{enumerate}

\item If we happen to get ${\cal M}_*$, then, by Observation 1, we can apply 
the Independence Property (see Section~\ref{indprop}), as we have a perfectly uniform $H$-PM. In this case Theorem~\ref{product} 
(see in Section~\ref{indprop}) provides that the new $A_i$ sets obtained in  Step 7 will 
remain ideal, i.e., independent and uniformly random elements  of $H$. Consistently with our introduced notation, let these ideal 
updated values for the next round be denoted by $A_*^{k+1}=(A_{*1}^{k+1},\ldots,A_{*n}^{k+1})$.
That is, each $A_{*i}^{k+1}$ is perfectly uniform over ${\cal I}(G_{k+1})$, and they are independent of each other.

\item If we happen to get $M_\psi$, then we obtain some other sets in Step 7, say $B^{k+1}=(B_1^{k+1},\ldots,B_n^{k+1})$. 
They may possibly be neither independent of each other, nor uniform over ${\cal I}(G_{k+1})$.
By the mixture (\ref{mixD}), however, this ``bad case" happens only with probability $\rho=2^{-Rm}$, which is a small enough value, 
such that we will be able to make  its effect negligible. 

\end{enumerate}
If $A^{k+1}=(A^{k+1}_1,\ldots,A^{k+1}_n)$ denotes the (possibly not ideal) system of updated values for the next round, obtained in Step 7, 
then by the above considerations we can represent the mixture (\ref{mixD})  in 
the following way. Let $\xi_k$ be a 0--1-valued independent random variable with $\Pr(\xi_k=1)=1-\rho$ and $\Pr(\xi_k=0)=\rho$.
(an independent coin flip with a biased coin). Then
\begin{eqnarray} \label{mixA}
A^{k+1} = \left\{
    \begin{array}{cl}
      A_*^{k+1}  & \;\;\mbox{ if  }\;\;\;\; \xi_k=1  \\[1.5mm]
      B^{k+1}  & \;\;\mbox{  if  }\;\;\;\;  \xi_k=0.
    \end{array} \right.
\end{eqnarray}
Now let $\beta_w$ denote the distribution of the output of the algorithm in the well behaved case, and let $Z$ denote the output itself. 
It is generated in Step 8, as the 
value of the set $A_1$, after all rounds are completed, i.e., when $k=m$. 
Let $E$ denote the "bad event`` that we happen to draw $\xi_k=0$ at least once in (\ref{mixA}), within the current repetition.
Since each $\xi_k$ is modeled as an independent biased coin flip, and $\Pr(\xi_k=1)=1-\rho$, therefore,
the probability that we always get $\xi_k=1$ is $\left(1-\rho\right)^{m}$. The complement of this event is the ``bad event" $E$, giving us
\begin{equation}\label{1-eps}
\Pr(E) \= 1- \left(1-\rho\right)^{m+1}.
\end{equation}
Consider now the ideal situation, when $\xi_1=\ldots=\xi_m=1$, leading to a uniform output. This ideal case happens 
with probability
$$
\Pr(\xi_1=\ldots=\xi_m=1) \= \left(1-\rho\right)^{m},
$$
and it generates an  output with distribution $U$, which  is
the uniform distribution over ${\cal I}(G)$.
Further, let $\beta_w^E$ denote the output distribution $\beta_w$, conditioned on $E$, that is, $\beta_w^E(x)=\Pr(Z=x\;|\;E)$. This represents the 
non-ideal case when we deviate from the ideal uniform distribution, i.e., we happen to draw $\xi_k=0$ at least once in (\ref{mixA}).
Then the overall output distribution is the mixture of the ideal and non-ideal cases, weighted with the corresponding probabilities:
$$
\beta_w \= (1-\rho)^{m}U + \left(1- (1-\rho)^{m}\right)\beta_w^E.
$$
By the Mixture Representation Property of the separation distance (see (iv) in Lemma~\ref{lemma1}), this yields
\begin{equation}\label{beta}
s(\beta_w,U) \leq 1-(1-\rho)^{m} \=1-(1-2^{-Rm})^{m}
\end{equation}
where in the right-hand side we used $\rho=2^{-Rm}$ from (\ref{sdu}). To get a simpler upper bound on $s(\beta_w,U)$, 
we can obtain from the well known Bernoulli inequality that $(1-2^{-Rm})^{m}\geq 1-m2^{-Rm}$ holds. 
Applying it to (\ref{beta}), the rightmost side becomes upper bounded 
by $1-(1-m2^{-Rm})=m2^{-Rm}$, resulting in 
\begin{equation}\label{beta2}
s(\beta_w,U) \leq m2^{-Rm}.
\end{equation}
Since $m$ is the number of edges in an $N$-vertex graph, therefore,  we have $m\leq N^2$. Further,  if $G$ 
has no isolated vertex, then $m\geq N/2$.
({\em Note:} if there are isolated vertices, we can handle them in  the following way: remove the isolated vertices;
run the algorithm on the remaining graph, and then add each isolated vertex to the result independently with probability 1/2.) 
Thus, we can conclude from (\ref{beta2}):
\begin{equation}\label{beta2b}
s(\beta_w,U) \leq m2^{-m}  \leqq N^2 \, 2^{-RN/2}.
\end{equation}
This gives us an exponentially small upper bound on the separation distance between the  output distribution in the well-behaved case ($\beta_w$) and 
the ideal uniform distribution $U$.

Let us now look into the deviant case, which means that the matrix $X$ does not have an $H$-PM in some round in {\em every} repetition.
Then the algorithm  eventually and halts with output $\{v_0\}$ in Step 5.
Let us now analyze the impact of the deviant case on the output distribution.

Let $p(X)$ denote the probability that a considered random  matrix $X$ does not have an $H$-PM.
We are going to show in Theorem~\ref{pm2} (see Section~\ref{probpm}) that this probability is exponentially small. Specifically, 
there exists a constant $N_0$, such that 
\begin{equation}\label{pX}
p(X)\leq {\rm e}^{-N}
\end{equation}
holds, whenever $N\geq N_0$. This requires the choice of $n=2N^2$, which is set in Step 1. The value of the constant $N_0$ is not particularly large,
$N_0=21$ suffices (see Theorem~\ref{pm2} in Section~\ref{probpm})\footnote{Note that (\ref{pX}) is significantly harder to prove than it seems, due to the 
dependencies among the matrix entries. We dedicate a section plus an 8-page appendix to the  proof. We suspect that it is not the simplest possible proof,
but at the time of submission we did not have a shorter one.}. 

Let $p_0$ denote the probability that failure occurs in a given repetition, that is, at least in one of the rounds 
the random matrix $X$ generated in the round does not have an $H$-PM, in the considered  repetition. 
Taking into account that we check the existence of an $H$-PM
at most $m$ times in each repetition (once in each round, except the initialization), we obtain 
from (\ref{pX}) that the deviant case can be encountered in a given repetition with probability at most
\begin{equation}\label{p0}
p_0\leqq m{\rm e}^{-N} \leqq N^2{\rm e}^{-N}
\end{equation}
whenever $N\geq N_0$. 

Let $P_d$ denote the probability that the deviant case occurs, that is, a failure is encountered in each of the $R$ repetitions. Taking
(\ref{p0}) into account, as well as the independence of the repetitions, we can write
\begin{equation}\label{P_d}
P_d \=  p_0^R \leqq N^{2R} {\rm e}^{-RN}
\end{equation}
whenever $N\geq N_0$.

Let us denote the output distribution of the Algorithm in the deviant case by $\beta_d$ 
(it is concentrated on the single vertex $v_0$, see Step 5). 
Further, let $\beta$ be the overall output distribution, regardless of whether
the case is deviant or well-behaved. Then $\beta$ is the mixture of $\beta_w$ and $\beta_d$, with weights representing the 
probabilities that the two cases occur. Let these probabilities be $P_w$ and $P_d$. Then $P_w+P_d=1$ and we can write:
$$
\beta \= P_w \beta_w + P_d \beta_d \= (1-P_d)\beta_w + P_d\beta_d.
$$
By the Mixture Representation Property of the separation distance (see (iv) in Lemma~\ref{lemma1}), this is equivalent to 
$s(\beta, \beta_w) \leq P_d$, so together with (\ref{P_d}) we get
\begin{equation}\label{beta3}
s(\beta, \beta_w) \leqq N^{2R} {\rm e}^{-RN}
\end{equation}
for $N\geq N_0$.
From the Triangle Inequality Property of the separation distance (see (iii) in Lemma~\ref{lemma1}), we obtain 
$$
s(\beta, U) \leqq s(\beta, \beta_w) + s(\beta_w,U).
$$
The above inequality  yields, using  (\ref{beta3}) and (\ref{beta2b}), that 
\begin{equation}\label{beta4}
s(\beta, U) \leqq  N^{2R} {\rm e}^{-RN} + N^2 \, 2^{-RN/2}.
\end{equation}
holds, whenever $N\geq N_0$. This can be bounded from above as 
\begin{equation}\label{beta5}
s(\beta, U) \leqq 2N^{2R}\, 2^{-RN/2}
\end{equation}
for $N\geq N_0$. 
We want to choose $R$ to satisfy  $s(\beta, U) \leq \vareps$, which can be achieved if
\begin{equation}\label{R}
2N^{2R} 2^{-RN/2} \leqq \vareps.
\end{equation}
One can check with elementary calculations that (\ref{R}) is  satisfied with
\begin{equation}\label{R1}
R \= \left\lceil \log_2 \frac{2}{\vareps} \right\rceil
\end{equation}
if $N\geq 17$. We indeed set the value (\ref{R1}) for $R$ in Step 1. We still have to assume $N\geq N_0$, so that (\ref{beta5}) continues to hold.
Thus, we obtain that with the choice of $R=\lceil \log_2 (2/\vareps) \rceil $, the bound $s(\beta, U) \leq \vareps$ will hold for $N\geq N_0=21$.
Taking into account that the separation distance is always an upper bound on the total variation distance (see (v) in Lemma~\ref{lemma1}), we 
eventually obtain 
\begin{equation}\label{beta6}
d_{TV}(\beta, U) \leqq  \vareps 
\end{equation}
whenever $N\geq N_0$, as desired.

Finally, considering the running time, observe that the only part of Algorithm 1 that is not clearly doable in polynomial time within any given 
repetition is Step 6: finding the approximately uniform $H$-PM. 
(Note that checking the existence of a perfect matching in Step 5 can be accomplished by a single call to a maximum matching 
algorithm, so in this sense it is clearly doable in polynomial time in $N$.)

In Step 6, the finding of the approximately uniform $H$-PM is accomplished by Algorithm 3, building on Theorem~\ref{JSV},
so it is also doable in probabilistic polynomial time. 
Note that the error parameter $\delta$ of Theorem~\ref{JSV} also plays a role in the running time of Algorithm 3.
Specifically, its running time depends polynomially on $\log \delta^{-1}$.
Due to our choice of $\delta=2^{-n-Rm-1}$, we have that $\delta^{-1}$ grows exponentially with $n+Rm=2N^2+Rm$. However, 
$\log \delta^{-1}$ grows only  linearly with $n+Rm$,
so it results in polynomially bounded running time in $N$ and $R$. Taking into account 
the choice of $R=\lceil \log_2 (2/\vareps) \rceil $, 
we obtain that Algorithm 3 has an overall probabilistic polynomial running time in $N$ and $\log \varepsilon^{-1}$.
Finally, considering that there are $R$ repetitions, and Algorithm 3 is run at most $m$ times in each one,
we conclude that Algorithm 1 has an overall probabilistic polynomial running time in $N$ and $\log \varepsilon^{-1}$.

\hspace*{10mm} \hfill $\spadesuit$

\subsection{Probability of Having a Perfect Matching in a Random Matrix}
\label{probpm}

Recall that in Step 5 of Algorithm 1 we check whether the random matrix $X$ has an $H$-PM.
In this section we prove that not having an $H$-PM happens only with  probability that is exponentially small as a function of  
the matrix size. Due to the allowed dependencies among the matrix entries, however, this is harder to prove than one would expect.
The result of this analysis is summarized in the following theorem:

\begin{theorem}\label{pm2}
Let $X$ be the $n\times n$ sized random matrix created in {\rm Steps 3} and {\rm 4} of Algorithm 1, and set $n=2N^2$. 
Denote by $p(X)$ the probability that $X$ does not have an $H$-PM.
Then for $N\geq 21$ the value of $p(X)$ is bounded as
$$
p(X)\leq {\rm e}^{-N}.
$$
\end{theorem}

\noindent
{\bf Proof.} See Appendix B.

\section{FPRAS for the Approximate Counting of Independent Sets}
\label{FPRAS}

Having presented a fully polynomial almost uniform sampler of independent sets in Section~\ref{almost}, we now use known results from the literature 
to transform it into a Fully Polynomial Randomized Approximation Scheme (FPRAS) for {\em counting} the independent sets in bounded degree graphs. 

Let us first define what exactly is that we are looking for. We specify the definition of FPRAS for the case of independent sets in graphs, because that is 
what we consider here, but the definition can be directly modified to other counting problems, as well.

\begin{definition}\label{FPRASdef} {\bf (FPRAS for counting independent sets)}
Let $G$ be a graph with $n=|V(G)|$ and let $I(G)=|{\cal I}(G)|$ denote the number of independent sets in $G$. 
Further, let $\vareps_0>0$ be a tolerance parameter. We say that a (randomized) algorithm is a 
{\em Fully Polynomial Approximation Scheme (FPRAS)} for $I(G)$, if it satisfies the following requirements:
\begin{enumerate}

\item[\rm (i)] Upon input $G,\vareps_0$, the algorithm outputs a number $\widetilde I$, such that 
\begin{equation}\label{FPRAS1}
\Pr\lor\, (1-\varepsilon_0)I(G)\leq \widetilde I \leq (1+\varepsilon_0)I(G)v) \ler \geqq \frac{3}{4}
\end{equation}
where the probability is meant solely with respect to the internal randomization of the algorithm (the input is not random).

\item[\rm (ii)] The running time of the algorithm is bounded by a polynomial of $n$ and  $1/\vareps_0$.

\end{enumerate}
\end{definition}

\noindent {\bf Remark.} The value 3/4 in (\ref{FPRAS1}) is somewhat arbitrary, it could be replaced by any constant in $(1/2,1)$, so that the 
definition still remains equivalent with the original. In fact, it still equivalent with the following, seemingly more demanding definition:
given $G, \vareps_0$, and, in addition, also a confidence parameter $\rho$, output a number $\widetilde I$, such that the algorithm runs in probabilistic 
polynomial time in $n,1/\vareps_0, \log (1/\rho)$, and it satisfies
\begin{equation}\label{1-eps2}
\Pr\left( (1-\varepsilon_0)I(G)\leq \widetilde I \leq (1+\varepsilon_0)I(G)v) \right) \geqq 1-\rho.
\end{equation}
This, apparently more demanding definition explains why the scheme is called {\em fully polynomial:} the running time does not become exponential 
even for exponentially small $\rho$, as the running time is polynomial in $\log (1/\rho)$,
rather than in $1/\rho$. It turns out, however, that 
it is sufficient to satisfy (\ref{1-eps2})  with a {\em constant}  $\rho=1/4$, which gives rise to the simpler formulation (\ref{FPRAS1}). 
The reason is that 
once we have an algorithm for the simpler version (\ref{FPRAS1}), we can boost the confidence parameter
to an arbitrary given $\rho\in (0, 1/2)$.
It can be achieved this way: given an arbitrary $\rho\in (0,1/2)$, run the algorithm
$k=\lceil 16\ln (2/\rho)\rceil$ times with independent random bits, each time with confidence parameter $\rho=1/4$, 
and then take the median of the obtained estimates.
The median will satisfy (\ref{FPRAS1}) with the desired value of $\rho$, while the running time increases only by a factor of 
$O(\log(1/\rho))$, so it is  polynomial in $n, 1/\vareps_0,\log (1/\rho)$. This claim appears to be part of folklore.
For a proof see, e.g., Vigoda \cite{vigoda}; it is a straightforward application of the Chernoff bound. 
Because of this equivalence, usually the simpler definition (\ref{FPRAS1}) is used in the literature, we also follow this tradition.

\medskip
Our next task is to transform our fully polynomial almost uniform {\em sampler} of independent sets into an FPRAS for {\em counting} independent sets.
As already noted in the Introduction, it has been well known for decades that the approximately  uniform
random  sampling  of  combinatorial  structures  and  the  approximate
counting of these  structures are closely  related, see e.g.,  Jerrum and Sinclair \cite{jerrum1},
Jerrum, Valiant,   and   Vazirani   \cite{jerrum2},   Sinclair   and    Jerrum
\cite{sinclair2}, Vigoda \cite{vigoda}, and numerous others. 

The typical method for deriving the counting algorithm from an almost uniform sampler is to express the counting task in the
form of a telescopic product of fractions, and then estimate the factors  via random samples. 
This is carried out in the literature for a good number of tasks, such as counting  matchings, graph colorings, knapsack solutions, various counting problems related to statistical physics, etc. Interestingly, however, it is not so easy to come by a definite reference  where 
the details are spelled out for independent sets, {\em and} the claim is also stated as a precise theorem, 
so that we can just plug our results into it, without reproducing lengthy proofs.
 Eventually, we found the online lecture note of Dyer, Frieze, and Jerrum \cite{mixingbook}. To state the result from \cite{mixingbook}, we first need to 
 repeat their definition of a {\em Good Sampler} (see in \cite{mixingbook}, Section 1.2.1.), adapted to our notations:

\begin{definition}
{\bf (Good Sampler)}
A {\em Good Sampler} for ${\cal I}(G)$ is a randomized algorithm that takes input as graph $G$ and a tolerance $\delta>0$, 
and produces an independent set $I$ (a random variable) such that  the probability distribution of $I$ is within total variation distance
$\delta$ of the uniform distribution on ${\cal I}(G)$. 
\end{definition}
Observe that the running time is not specified directly in the definition of a Good Sampler, but the following theorem deals with it.
\begin{theorem}\label{DFJ}
{\rm (Dyer, Frieze, and Jerrum \cite{mixingbook}, Theorem 1.2.1)}
Suppose we have a a Good Sampler for the independent sets of a graph, which works for graphs $G$ with
maximum degree bounded by $\Delta$ and suppose the sampler has time complexity $T(N,\delta)$, where $N$ is the number of vertices 
in $G$ and $\delta$ is the allowed deviation from uniformity in the sampling distribution\footnote{Note that the parameter $\delta$ was 
denoted by $\vareps$ in our Definition~\ref{eaus}.}. Then we may construct an FPRAS for 
the number of independent sets of a graph, which works for graphs $G$ with maximum degree bounded by $\Delta$, and which has 
time complexity
$$
O\left( \frac{m^2}{\varepsilon_0}\, T\left(N, \frac{\varepsilon_0}{6m} \right) \right),
$$
where $N,m$ are the number of vertices and edges in $G$, respectively, and $\varepsilon_0$ is the specified error bound in (\ref{FPRAS1}).

\end{theorem}
Based on Theorem~\ref{DFJ}, we can now claim the existence of an FPRAS for counting the independent sets in bounded degree graphs:

\begin{theorem} \label{FPRASthm}
{\bf (FPRAS for independent sets in bounded degree graphs)}
Let $\Delta$ be an arbitrary fixed positive integer that serves as an upper bound on the maximum degree of the considered graphs. 
There exists an FPRAS for counting 
the number of independent sets of graphs, which works for any graph $G$ with maximum degree bounded by $\Delta$, and which has 
time complexity
$$
O\left( \frac{m^2}{\varepsilon_0} T\left(N, \frac{\varepsilon_0}{6m} \right) \right),
$$
where $N,m$ is the number of edges in $G$, respectively, and $\varepsilon_0$ is the specified error bound in (\ref{FPRAS1}).

\end{theorem}

\noindent {\bf Proof.} We can use our Algorithm 1 (see Section~\ref{almost}) for the Good Sampler in Theorem~\ref{DFJ}. Observe that our algorithm 
actually provides more than what is needed: (1) it does not require an upper bound on the maximum degree for sampling; and  (2) its running time is 
polynomial in $N$ and $\log(1/\delta)$, even though it would be enough to have here $1/\delta$, instead of $\log(1/\delta)$. 
(Note that $\delta$ corresponds to what was denoted by $\vareps$ in Definition~\ref{eaus}.)

Either way, using our Algorithm 1 as a Good Sampler, Theorem~\ref{DFJ} implies that for any degree bound $\Delta$, an FPRAS exists for counting the independent sets in graphs with maximum degree at most $\Delta$. 

{\em Note:} The fact that our Algorithm 1 and Theorem~\ref{main} do not use a degree bound does not imply that the FPRAS exists without assuming a degree bound, as the proof of Theorem~\ref{DFJ} uses the degree bound. However, it can be an arbitrary constant, as Algorithm 1 
works without degree bound. 

\hfill $\spadesuit$

\section{NP vs.\ RP}
\label{NPvsRP}

Finally, we are ready to state our main result. Its proof may appear deceptively short. Keep in mind, however, that the theorems on which it is built 
are far from simple.

\begin{theorem}
  $\;{\bf NP}={\bf RP}$.
\end{theorem}

\noindent {\bf Proof.} By  Theorem~\ref{FPRASthm}, 
an FPRAS exists for  approximately counting
the independent sets in bounded degree graphs, with any constant degree bound $\Delta$. 
This is known  to imply ${\bf NP}={\bf RP}$, with at least three different
proofs  for  it  in  the  literature.  Specifically,  Luby  and Vigoda
\cite{luby}  prove  in  1997,  building  on  an  approach presented by
Sinclair \cite{sinclair}\footnote{Sinclair \cite{sinclair}  attributes
the idea of  boosting the number  of solutions to  Jerrum, Valiant and
Vazirani \cite{jerrum2}.}, that there is a constant $\Delta$, such that the
existence  of  the  FPRAS  for  graphs with maximum degree  $\leq \Delta$ implies ${\bf
NP}={\bf RP}$. The unspecified constant $\Delta$ is brought down to  $\Delta=25$
by  Dyer,  Frieze,  and  Jerrum  \cite{dyer}  in  2002.  Finally,  Sly
\cite{sly} further  reduces the  constant to  $\Delta=6$ in  2010. All  the
three papers prove, with different  methods, that the existence of  an
FPRAS  for  approximately  counting  independent  sets  in graphs with
the respective constant degree bound $\Delta$ implies ${\bf NP}={\bf RP}$.

\hfill $\spadesuit$

\section{Some Notes on Other Complexity Classes}
\label{other}

The intriguing  question  still remains  whether  ${\bf RP}$ can be
derandomized or not. In  any case, if our  result is correct, then now  the
two commonly accepted conjectures  ``${\bf P}\neq  {\bf NP}$"  and ``${\bf  P}={\bf RP}$"
now contradict to  each other. Our opinion, contrary to the majority view, is that ${\bf P}\neq{\bf RP}$, i.e., 
complete derandomization is not possible. If this is indeed true, then we have
${\bf P}\neq{\bf RP}={\bf NP}$. This would also give some hint why the {\bf P} vs.\ {\bf NP} problem has been so difficult:
\begin{itemize}

\item If our result is correct, then {\bf NP} is {\em almost} equal to {\bf P}, the {\em only} barrier is the need for random bits. Therefore, separating
{\bf P} and {\bf NP} is hard, as they are {\em very close} to each other. Yet, if complete derandomization is not possible,  then
they are still different, so collapsing them is impossible. 

\item If ${\bf P}\neq{\bf RP}={\bf NP}$ is actually the case, then, in a sense, it is just the opposite of what has been widely believed, which is 
${\bf P}={\bf RP}\neq {\bf NP}$. While the statements ${\bf P}\neq{\bf RP}={\bf NP}$ and ${\bf P}={\bf RP}\neq {\bf NP}$ are not formal negations of each other, they intuitively point in opposite directions. Thus, one could possibly argue that previous research, no matter how involved, 
could not reach the goal, because it pursued the wrong direction. 

\end{itemize}

Some consequences of ${\bf  NP}={\bf RP}$ immediately follow  from existing
results.  For  example,  it  is  well  known  (proved by Zachos
\cite{zachos}),  that  ${\bf  NP}\subseteq  {\bf  BPP}$  implies   the
collapse  of  the  entire  Polynomial  Hierarchy  to {\bf BPP}, so our
result (if correct) implies ${\bf PH}={\bf BPP}$.

Another interesting fact is the issue of {\em average case complexity}
of $\bf NP$, in the  sense of Levin \cite{levin2}, assuming  the input
is chosen  randomly from  any reasonable  (polynomial-time computable)
distribution. According to current  knowledge, ${\bf P}\neq {\bf  NP}$
alone is not  known to rule  out that all  problems in {\bf  NP} could
still be solved in deterministic polynomial time {\em on the average}, over random inputs.
In other words, in case of ${\bf P}\neq {\bf NP}$, we could still have
that  {\bf  NP}  is  ``easy  on average" for deterministic algorithms.
However, Buhrman, Fortnow and Pavan \cite{buhrman} prove that if  {\bf
NP} is easy on average then ${\bf P}={\bf BPP}$. Since, by our result,
the latter would now imply ${\bf P}={\bf NP}$, therefore, whether $\bf
NP$ is hard in the worst case becomes {\em equivalent} to whether it is hard
in the average case.

It is also interesting  to look into that  if ${\bf P}\neq {\bf  NP}$,
then, combined with the  obtained collapse ${\bf NP}={\bf  RP}$, could
we conclude  something new  about complexity  class {\em separations}?
Here is a  consequence that we  could call ``disjunctive  separation,"
(conditional on ${\bf P}\neq {\bf NP}$) because it proves that {\em at
least one} of  two major separations  would follow. (The  same has not
been known to follow  from ${\bf P}\neq {\bf  NP}$ alone.) We use  the
common   notations   ${\bf   E}=   DTIME(2^{O(n)})$   and  ${\bf  NE}=
NTIME(2^{O(n)})$.

\begin{theorem}
If ${\bf P}\neq {\bf NP}$, then at least one of the following holds:

{\rm (1)}
${\bf NP}\neq \,co{\rm  -}{\bf NP}$

{\rm (2)}
  ${\bf E}\neq \,{\bf NE}$.
  \end{theorem}

\noindent {\bf Proof.} By our result ${\bf NP}={\bf RP}$ holds, which implies ${\bf PH}={\bf  BPP}$
(see  Zachos  \cite{zachos}).
Let  {\bf TALLY}  be the  set of all tally (unary) languages. Let us consider ${\bf TALLY}\cap {\bf PH}={\bf
TALLY}\cap  {\bf  BPP}$,  which  is  nonempty, as there are tally languages in {\bf PH}.  By  a  result of Babai,
Fortnow, Nisan and  Wigderson \cite{babai} if  all tally languages  in
{\bf PH} are decidable in polynomial time, then ${\bf P}={\bf  BPP}$.
Since ${\bf  P}\neq {\bf  NP}$ is  assumed, therefore,  by our result,
${\bf  P}={\bf  BPP}$  is  impossible.  Consequently, it must be that
${\bf TALLY}\cap ({\bf BPP}-{\bf P})\neq \emptyset$. Then {\bf NP} = {\bf RP} $\subseteq$ {\bf BPP}
 implies that  at
least  one   of  ${\bf   TALLY}\cap({\bf  NP}-{\bf   P})$  and   ${\bf
TALLY}\cap({\bf  BPP}-{\bf  NP})$  is  nonempty.  If  there is a tally
language in ${\bf NP}-{\bf  P}$ then it is  known to imply  ${\bf
E}\neq {\bf NE}$  (see Book \cite{book}).  If ${\bf BPP}-{\bf  NP}\neq
\emptyset$, then ${\bf NP}\neq  \,co{\rm -}{\bf NP}$ must  hold, since
otherwise  {\bf  BPP}  would  collapse  into  {\bf  ZPP}, due to ${\bf
NP}={\bf RP}$. If, however, {\bf  BPP}   collapses  into  {\bf  ZPP}, then 
we get the   contradiction  $\emptyset\neq {\bf BPP}-{\bf  NP}={\bf ZPP}-{\bf
RP}= \emptyset$.

\hfill $\spadesuit$

\section{Conclusion}
\label{conc}

We (claim to) have  proved ${\bf  NP}={\bf RP}$.  The most important
task is now the validation of the correctness by the expert community.

As a concluding remark, let us mention that out result currently proves the {\em theoretical} possibility 
for solving {\bf NP}-complete problems efficiently, i.e., in probabilistic polynomial time. 
This does not mean, however, that out method immediately yields a {\em practically fast} solution. In fact, it is still 
very slow, despite its polynomial running time. Nevertheless, we believe that once the possibility of an efficient solution is established, progress 
will eventually lead to a practical solution.

\medskip

\noindent
{\Large\bf Appendix A:  Properties of the Separation Distance}

\medskip
\medskip

\noindent {\bf Proof of Lemma~\ref{lemma1}.} Statements (i) and (ii)
({\bf\em Normalized Value} and
{\bf\em Separation}) directly  follow from the  definition. For the
{\bf\em
Triangle Inequality}   in      (iii),     let      $s(\alpha,\beta)=\varepsilon_1$,
$s(\beta,\gamma)=\varepsilon_2$.  Then  $\alpha\geq  (1-\varepsilon_1)\beta$
and    $\beta\geq    (1-\varepsilon_2)\gamma$,    implying    $\alpha\geq
(1-\varepsilon_1)(1-\varepsilon_2)\gamma                                \geq
(1-\varepsilon_1-\varepsilon_2)\gamma$,  which  yields $s(\alpha,\gamma)\leq
\varepsilon_1+\varepsilon_2 =s(\alpha,\beta)+s(\beta, \gamma)$.

\medskip

Considering the {\bf\em Mixture  Representation} (iv), we first observe that 
if $\varepsilon=0$, then
$\alpha=\beta$ holds and $\gamma$ can be arbitrary. If $\varepsilon = 1$,
then $s(\alpha,\beta)\leq 1$ is always true and $\gamma=\alpha$ is the 
unique distribution satisfying $\alpha=(1-\varepsilon)\beta+\varepsilon\gamma$.  
Thus,  the  only  nontrivial  case is $0<\varepsilon<1$, which we
treat below.

1. Assume
$s(\alpha,\beta)\leq \varepsilon$.
Let us define a real valued function $\gamma$ on $S$ by
\begin{equation}\label{gamma}
\gamma = \frac{1}{\varepsilon}
\alpha+\left(1-\frac{1}{\varepsilon}\right)\beta.
\end{equation}
By definition,
$\alpha \geq (1-s(\alpha,\beta))\beta$ always holds. Together with
$s(\alpha,\beta)\leq \varepsilon$ they imply
$$\frac{1}{\varepsilon}\alpha \geq
\frac{1-s(\alpha,\beta}{\varepsilon}\beta =
\frac{1}{\varepsilon}\beta - \frac{s(\alpha,\beta}\varepsilon\beta
\geq \frac{1}\varepsilon\beta - \beta.$$
Now using the definition of $\gamma$ we obtain
$$\gamma = \frac{1}{\varepsilon} \alpha
+\left(1-\frac{1}{\varepsilon}\right)\beta
\geq
 \frac{1}\varepsilon\beta - \beta +\beta - \frac{1}{\varepsilon}\beta
= 0.$$
Thus, $\gamma$ is nonnegative. Moreover,
$$\sum_{x\in S} \gamma(x) =
\frac{1}{\varepsilon} \underbrace{\sum_{x\in S}\alpha(x}_{=1}
\; + \;
\left(1-\frac{1}{\varepsilon}\right)
\underbrace{\sum_{x\in S}\beta(x}_{=1}
= 1,
$$
so $\gamma$ is indeed a probability distribution. Expressing $\alpha$
from (\ref{gamma}), we get
$\alpha = (1-\varepsilon) \beta + \varepsilon \gamma$. It is also clear that whenever $\varepsilon>0$, 
the equation $\alpha = (1-\varepsilon) \beta + \varepsilon \gamma$ uniquely determines $\gamma$.

\medskip

2. Conversely, suppose that a
probability distribution $\gamma$ exists on $S$ with
$\alpha = (1-\varepsilon) \beta + \varepsilon \gamma.$ Expressing $\gamma$
gives
$\gamma = (1/\varepsilon)\alpha
+ \left(1-1/\varepsilon\right)\beta.$
Assume indirectly that $s(\alpha,\beta)>\varepsilon$. Then there is an
$x\in S$ with $\alpha(x)<(1-\varepsilon)\beta(x).$ It implies
$$\gamma(x) =
\frac{1}{\varepsilon} \alpha(x) +
\left(1-\frac{1}{\varepsilon}\right)\beta(x) <
\frac{1}{\varepsilon} (1-\varepsilon)\beta(x)
+ \left(1-\frac{1}{\varepsilon}\right)\beta(x)
= 0,$$
yielding $\gamma(x)<0,$ which is impossible, since $\gamma$ is a
probability distribution. Thus, $s(\alpha,\beta)\leq \varepsilon$ must
hold.

\medskip

The proof of property (v) {\bf\em Connection to total variation distance} can be found in the paper of Aldous and Diaconis
\cite{aldous-diaconis}.

\medskip

Finally, to prove property (vi) {\bf\em Relationship to total variation distance for uniform distribution}, set $d_{TV}(\alpha, U)=\delta$. Since
we have
$$
d_{TV}(\alpha, U) \= \frac{1}{2}\sum_{x\in S} |\alpha(x)-U(x)| \=
\frac{1}{2}\sum_{x\in S} \left|\alpha(x)-\frac{1}{|S|}\right|,
$$
therefore,
$$
\frac{1}{2} \left|\alpha(x)-\frac{1}{|S|}\right| \leqq \delta
$$
must hold for any $x\in S$. Rearranging the above inequality, we get
$$
\largabs |S|\alpha(x) - 1 \largabs \leqq 2|S|\delta,
$$
which is equivalent to 
\begin{equation}\label{delta}
1-2|S|\delta \leqq |S|\alpha(x) \leqq 1+2|S|\delta.
\end{equation}
Using the formula (\ref{ald}) for the separation distance with $\beta=U$ yields
\begin{equation}\label{delta2}
s(\alpha,U)\=\max_x \left(1-\frac{\alpha(x)}{U(x)}\right) \=
\max_x \left(1-\frac{\alpha(x)}{1/|S|}\right) \=
1- \min_x (|S|\alpha(x)).
\end{equation}
Observe now that by (\ref{delta}) we have $|S|\alpha(x) \geq 1-2|S|\delta$ for every $x\in S$, which implies 
$$\min_x (|S|\alpha(x)) \geqq 1-2|S|\delta.$$
This inequality can be rearranged as $1-\min_x (|S|\alpha(x)) \leq 2|S|\delta$, which together with (\ref{delta2}) yields
$ s(\alpha,U) \leq 2|S|\delta$. Combining it $d_{TV}(\alpha, U)=\delta$, we get 
$$
s(\alpha,U) \leqq 2|S|d_{TV}(\alpha, U),
$$
what we wanted to prove.

\hspace*{10mm} \hfill $\spadesuit$

\medskip\medskip

\noindent
{\Large\bf 

\noindent
Appendix B:  Probability of Having a Perfect 
Matching in a \\[1.5mm	] Random Matrix -- Proof of Theorem~\ref{pm2}}

\medskip\noindent
Let $X$ be the random matrix that we create in each round in Steps 3 and 4 of Algorithm 1, 
and let $B=X^H$ be the $H$-skeleton of $X$ (see Definition~\ref{H-skel}).
Set $H_1=\{1\}$. Then the 0--1 matrix $B=X^H$ has an $H_1$-PM if and only if $X$ has an $H$-PM.
For simplicity, we call an $H_1$-PM in the 0--1 matrix $B$ a perfect matching, as it corresponds to a traditional perfect matching in the 
bipartite graph $G_B$. Thus, it is enough for us to study the probability that a random 0--1 matrix has a perfect matching.

Unfortunately, the situation is made more complicated by the fact that the entries of the random matrix $B$ are not independent. 
Recall that $X$ arises in Steps 3 and 4 in Algorithm 1 by running $n$ 
 realizations of the BIDC Markov chain independently, and filling each one into a row of the matrix. As a result, 
the entries within a row belong to the same realization, so they are generally not independent. Then $B$ is obtained by taking the 
$H$-skeleton of $X$ (see Definition~\ref{H-skel}). To capture the kind of non-independent randomness that the matrix $B=X^H$ exhibits, let us introduce some
definitions below.

\begin{definition} \label{preced} {\bf (Preceding entry)} 
Let $B=[b_{ij}]$ be a matrix of size $n\times n$. We say that the  index pair $(i,j)$ precedes another index pair $(k,\ell)$, 
denoted by $(i,j)\prec (k,\ell)$, if either $i<k$, or $i=k$ and $j<\ell$. Then we also say that the entry $b_{ij}$ precedes the
entry $b_{k\ell}$, denoted by $b_{ij} \prec b_{k\ell}$.
\end{definition}
In words, for any entry, its preceding entries are those that are located either in a row that is higher up, or in the same row, but to the left
of the considered entry. 

\begin{definition} \label{defp-robmx} {\bf ($p$-robust random 0--1 matrix)} 
Let $Y=[\xi_{ij}]$ be a random 0--1 matrix, drawn from an arbitrary distribution, and $p\in (0,1)$.
We say that  $Y$ is {\em $p$-robust}, if $\forall i,j: \; \Pr(\xi_{i,j}=1)\geq p$, 
and this remains so even if we condition on arbitrary values of any
preceding matrix entries. Formally, if $\xi_{i_1j_1},\ldots,\xi_{i_mj_m}$ precede $\xi_{ij}$, then 
$$\Pr(\xi_{ij}=1\;|\; \xi_{i_1j_1}=a_{i_1j_1},\ldots,\xi_{i_mj_m}=a_{i_mj_m}) \geqq p$$
holds for every $i,j,$ and for every choice of the constants  $a_{i_1j_1},\ldots,a_{i_mj_m}  \in \{0,1\}$, whenever the condition has positive probability.  
\end{definition}
In words, the matrix is $p$-robust, if any entry takes the value 1 with probability at  least $p$, regardless of
fixing the values of any preceding entries. Note that it does not mean the considered entry is independent of the preceding ones, as 
the probability that it takes the value 1 may depend on them, just this probability never drops below $p$. 


Next we show that the matrix $B=X^H$ is indeed $p$-robust with an appropriate choice of $p$.

\begin{lemma}\label{p-roblemma} 
If $N\geq 4$, then the matrix $B=X^H$ is $p$-robust with $p=\frac{2}{N}$, where $N$ is the number of vertices 
in the input graph $G$ of Algorithm 1.
\end{lemma}

\noindent {\bf Proof.} Recall that the $n\times n$ matrix $X$ is created in Steps 3 and 4 of Algorithm 1 as repeated below. (Note that 
$n$ is the number of realizations we run, which an internal parameter of the algorithm. It is  not the same as $N$, which is the number of vertices in $G$.)

\begin{quote}

{\bf Step 3 (Running the Markov chain in round $k$)} 
Run $n$  realizations of the BIDC Markov chain independently in $G_k$ (for the description of BIDC see Section~\ref{BIDC}).
Start the $i^{th}$ realization from 
initial state $A_i, \; i=1,\ldots,n$. Run each realization for $n$ steps, and let $X_{i0}, X_{i1},\ldots,X_{in}$ denote the state 
sequence taken by the $i^{th}$ realization, where $X_{i0}=A_i$. 

{\bf Step 4 (Creating the matrix)} Collect the $n$ realizations in an $n\times n$-sized matrix $X=[X_{ij}]$, 
so that the  $i^{th}$ realization is filled into the $i^{th}$ row, 
$ i\in [n]$, starting the row with $X_{i1}$. Set $S={\cal I}(G_k)$ as the domain containing all matrix entries,  and set $H={\cal I}(G_{k+1})\subseteq S$.

\end{quote}
Thus,  if $X_{i0}, X_{i1},\ldots,X_{in}$ is the state 
sequence taken by the $i^{th}$ realization, then the $i^{\rm th}$ row of $X$ becomes $X_{i1},\ldots,X_{in}$.
Observe now that by  running the Markov chain realizations independently, the value of $X_{it}$ is determined solely 
by the value of $X_{i,t-1}$, plus the independently drawn state transition that selects the next state where the chain moves from 
the previous state $X_{i,t-1}$. 

{\em Note:} The reason for filling $X_{i1},\ldots,X_{in}$ into the $i^{\rm th}$ row of $X$, not including 
$X_{i0}$, is that this gives us a previous  state even for the first entry in the row. This way each matrix entry has a previous one. 
Note that this is not the same as a preceding entry, defined in Definition~\ref{preced}. According to that definition, the entry
in the upper left corner does not have a preceding entry.
 
For any entry $X_{it}$ of $X$, let us call the $H$-status of $X_{it}$ the indicator of whether $X_{it}$ falls in $H$ or not. This is the same as 
the corresponding entry of the $H$-skeleton matrix $B=X^H$ (see Definition~\ref{H-skel}). Further, let $h(i,t)$ denote the vector that lists 
the $H$-status of all preceding entries of $X_{it}$, except $X_{i,t-1}$. We refer to $h(i,t)$ as the {\em $H$-history} of $X_{it}$. 
This $H$-history may be empty, if there is no preceding entry other than $X_{i,t-1}$. When we are going to condition on an $H$ history $h(i,t)=h_0$, 
it means that $h_0$ is a 0--1 vector, which we allow to  be an empty vector with no components at all, to include the case of an empty $H$-history. 

To prove the $p$-robustness of $B=X^H$ it is enough to show that for any $i,t\in [n]$ it holds
that $X_{it}$ falls in $H$ at least with probability $p$, no matter which of its preceding entries fall in $H$, if any. 
This means, whether or not $X_{i,t-1}\in H$ and whatever the value of the $H$-history $h(i,t)$ is, 
we have that $X_{it}$ falls in $H$ at least with probability $p$.

Recall now that 
$H={\cal I}(G_{k+1})$. If the edge $e_{k+1}$ connects vertices $a,b$, then an independent set $F\in S={\cal I}(G_k)$ also 
satisfies $F\in H={\cal I}(G_{k+1})$ precisely when $\{a,b\}\not\subseteq F$, that is, $F$ avoids at least one of $a$ and $b$. 
(Note that $e_{k+1}$ is an edge in $G_{k+1}$, but not in $G_k$.)

Using the above considerations, we can argue that if $X_{i,t-1}\in H$, then  $X_{i,t-1}$ avoids at least one of $a$ and $b$, regardless of $h(i,t)$. 
Therefore, if the uniform random vertex $u$ in the BIDC transition rule (see Section~\ref{BIDC}) satisfies $u\notin \{a,b\}$, then 
$X_{it}$ will still avoid at least one of $a,b$, so $X_{it}$ will remain in $H$, regardless of $h(i,t)$. As $\Pr(u\notin \{a,b\})=1-\frac{2}{N}$ 
in an $N$-vertex graph, we obtain that for any possible value $h_0$ of the $H$-history the following holds:
\begin{equation}\label{inH1}
\Pr\lor X_{it}\in H\;|\; X_{i,t-1}\in H,\; h(i,t)=h_0 \ler \geqq 1-\frac{2}{N}.
\end{equation}
If $X_{i,t-1}\notin H$, then  $X_{i,t-1}$ must contain both $a$ and $b$. In this case $X_{it}\in H$ happens if $u\in \{a,b\}$, because then
$u$ eliminates either $a$ or $b$ from $X_{i,t-1}$, so $X_{it}$ will avoid at least one of $a,b$, which is equivalent to $X_{it}\in H$. 
This again holds for any $H$-history $h(i,t)=h_0$.
Using that $\Pr(u\in \{a,b\})=2/N$ in an $N$-vertex graph, we get
\begin{equation}\label{inH2}
\Pr\lor X_{it}\in H\;|\; X_{i,t-1}\notin H\; h(i,t)=h_0 \ler \geqq \frac{2}{N}.
\end{equation}
From (\ref{inH1})and (\ref{inH2}) we see that no matter which of the preceding entries of $X_{it}$ fall in $H$,
the entry $X_{it}$ itself falls in $H$ at least with probability
\begin{equation}\label{mini}
p \= \min \left\{1-\frac{2}{N},\; \frac{2}{N} \right\},
\end{equation}
which means that $B=X^H$ is $p$-robust with the above value of $p$.
If $N\geq 4$, then $1-\frac{2}{N}\geq \frac{2}{N}$, so the minimum in (\ref{mini}) is $\frac{2}{N}$.
Thus, we obtain that for $N\geq 4$ the matrix $B=X^H$ is $p$-robust with $p=\frac{2}{N}$, what we wanted to prove.

\hspace*{10mm} \hfill $\spadesuit$

The next result shows that a $p$-robust matrix has a high probability of having a perfect matching.

\begin{theorem}\label{pm} 
Let $Y=[\xi_{ij}]$ be a random 0--1 matrix of size $n\times n$, drawn from an arbitrary distribution, and let $p\in (0,1)$.
If $Y$ is $p$-robust {\rm (see Definition~\ref{defp-robmx}),} then 
\begin{equation}\label{prpm}
\Pr(Y \, \mbox{\rm has a perfect matching})\geq 1- (n+1)^3 (1-p)^{\lfloor n/2\rfloor}.
\end{equation}
\end{theorem}
{\em Remark.} The bound (\ref{prpm}) is not the tightest possible, but it provides a simple, explicit, non-asymptotic formula, 
which will be valuable for our purposes.

For the proof of Theorem~\ref{pm} we reformulate the concepts in terms of random graphs, because it is easier to visualize, and allows us to use some 
graph theoretic tools.

\begin{definition}\label{defp-randsgr}
{\bf ($p$-random subgraph)}
Let $G$ be a graph and $p\in (0,1)$. Create a random subgraph of $G$ by keeping each edge with probability $p$, and deleting it 
with probability $1-p$, doing it independently for each edge. We denote the arising random graph by $G(p)$, and call it a 
$p$-random subgraph of $G$.
\end{definition}

\begin{definition}\label{defp-robsgr}
{\bf ($p$-robust random subgraph)}
Let $G=(V,E)$ be a graph with $E=\{e_1,\ldots,e_m\}$, and let $p\in (0,1)$. 
Create a random subgraph of $G$ by randomly keeping/deleting edges of $G$, but not necessarily independently. 
We require, however, that the following holds: each edge $e_i$ is kept with probability at least $p$,
given the status (kept/deleted) of  $e_{i+1},\ldots,e_m$, whatever that status is, assuming it has positive joint probability. 
For $i=m$, when the list 
$e_{i+1},\ldots,e_m$ is empty, we require that $e_m$ is kept with probability at least $p$, unconditionally.
The random subgraph arising this way is called a {\em $p$-robust random subgraph} of $G$, and is denoted by $\widetilde G(p)$.
\end{definition}
{\em Remark.} Observe that the definitions directly imply the following: a random 0--1 matrix is $p$-robust if and only if it is the bi-adjacency 
matrix of a $p$-robust random subgraph of a complete balanced bipartite graph, assuming that the graph edges are numbered in the following way: 
if entries $a_{ij},a_{k\ell}$ of the matrix correspond to graph edges $e_q,e_r$, respectively, then $a_{ij}\prec a_{k\ell}$ implies $q>r$.
If this is the case, we say that the edges are numbered consistently with the matrix precedence relation. Note that this consistent numbering 
simply means the following: read the matrix row by row, from bottom row upward to top, and from right to left in each row. Then assign 
the edges $e_1,\ldots,e_m$ to the matrix entries in this order. If the matrix is $p$-robust, then this will be 
equivalent to the requirement that each edge is kept with probability $p$, no matter whether the edges to the left of it in the same row, or any edge
in a higher row is kept or not.  

\medskip

\noindent {\bf Monotone graph property.}
Let $\cal Q$ be a set of graphs. We use it to represent a graph property: a graph has property $\cal Q$ if and only if $G\in \cal Q$. Therefore, we
identify the property with $\cal Q$.
We say that a graph property $\cal Q$ is {\em monotone,} if it is closed with respect to adding new edges. That is, if $G\in \cal Q$ and $G\subseteq G'$,
then $G'\in \cal Q$. (The notation $G\subseteq G'$ means that the edge set of $G$ is a subset of the edge set of $G'$, while their vertex set is the same.)

Below we prove that for any graph $G$, and for any monotone graph property, the following holds: the probability that a $p$-random 
subgraph of $G$ has the property can only increase if we switch to a $p$-robust random subgraph. It is interesting to note that while this claim is intuitively 
not hard to believe, it takes some effort to prove it formally.

\begin{lemma}\label{p-robgr}
Let $\cal Q$ be a monotone graph property, $G$ be a graph and $p\in (0,1)$. Then the following holds:
$$\Pr(G(p)\in {\cal Q})\;\leq \; \Pr(\widetilde G(p)\in {\cal Q})$$
where $G(p)$ and $\widetilde G(p)$ are the $p$-random and $p$-robust random subgraphs of $G$, respectively, defined in {\em Definitions 
\ref{defp-randsgr}} and {\em \ref{defp-robsgr}.}
\end{lemma}

\noindent 
{\bf Proof.} 
We are going to generate  $\widetilde G(p)$ as the union of two random graphs, $G(p)$ and $G_2$, both on the same vertex set $V(G)$. 
$G(p)$ is the $p$-random subgraph of $G$, the other random graph $G_2$ will be defined later.
The union $G(p)\cup G_2$ is meant with the understanding that if the same edge occurs  in both graphs, then we merge 
them into a single edge. We plan to chose the edge probabilities in $G_2$,
such that $G(p)\cup G_2\sim \widetilde G(p)$, where the ``$\sim$" relation between random graphs means that they have the same 
distribution, i.e., they are statistically indistinguishable.
If this can be accomplished, then the claim will directly follow, 
since then a random graph distributed as $\widetilde G(p)$ can be obtained by adding edges to $G(p)$, which cannot destroy a monotone property,
once $G(p)$ has it.  This will imply the claim of the Lemma.

Let us introduce some notations. Let $e_1,\ldots,e_m$ denote the edges of $G$. 
For every $i\in [m]$, let $h_i$ be the indicator of the event that the edge $e_i$ is included in the $p$-robust random graph  $\widetilde G(p)$. 
We denote this event by $\{e_i\in \widetilde G(p)\}$. 
Further, let us use the abbreviation $h_i^m=(h_i,\ldots,h_m)$.  
For any $a=(a_1,\ldots,a_m)\in \{0,1\}^m$, the event 
$\{h_1^m=a\}$ means that $\widetilde G(p)$ takes a realization 
in which edge $e_i$ is included if and only $a_i=1$. Similarly, $\{h_i^m=a_i^m\}$ means $\{h_i=a_i,\ldots,h_m=a_m\}$,
where we use the abbreviation $a_i^m=(a_i,\ldots,a_m)$.  

Now let us generate the random graphs $G(p)$ and $G_2$, as follows. 
\begin{description}

\item[\hspace*{5mm} \rm Step 1.] 
Let $i=m$.

\item[\hspace*{5mm} \rm Step 2.] 
If $i=m$, then let $q_m=\Pr(h_m=1)$.
If $i<m$, then set $q_i=\Pr(h_i=1\;|\; h_{i+1}^m=a_{i+1}^m)$, where  
$a_{i+1}^m$ indicates the already generated edges of $G(p)\cup G_2$.

\item[\hspace*{5mm} \rm Step 3.]
Compute  
\begin{equation}\label{p'_i}
p'_i=\frac{p(1-q_i)}{1-p}.
\end{equation}

\item[\hspace*{5mm} \rm Step 4.] 
Put $e_i$ into $G(p)$ with probability $p$, and put $e_i$ into $G_2$ with probability $q_i-p'_i$.

\item[\hspace*{5mm} \rm Step 5.]
If $i>1$, then decrease $i$ by one, and go to Step 2; else {\sc halt.}

\end{description}

\noindent 
{\em Remark:} In Step 2 we do not specify a procedure for how to actually compute the value of $q_i$.  
This algorithm, however, is only needed for the proof, its {\em existence} is enough for us. 
Therefore, it is sufficient to {\em define} $q_i$, without providing a specific procedure for its computation. Also observe that the condition in 
$q_i=\Pr(h_i=1\;|\; h_{i+1}^m=a_{i+1}^m)$ occurs with positive probability, as the condition represents the already generated part. For $i=m$ 
there is no condition, but Definition~\ref{defp-robsgr} implies $q_m=\Pr(h_m=1)>0$.

First note that the value $q_i-p'_i$ in Step 4 can indeed be used as a probability. Clearly, $q_i-p'_i\leq 1$ holds, 
as   $q_i$ is a probability and $p'_i\geq 0$. To show $q_i-p'_i\geq 0$, observe that 
$$ 
p'_i \=\frac{p(1-q_i)}{1-p} \leqq q_i,
$$
since the inequality can be rearranged into $p(1-q_i)\leq q_i(1-p)$, which simplifies to $p\leq q_i$. The latter is indeed true, due to 
$q_i=\Pr(h_i=1\;|\; h_{i+1}^m=a_{i+1}^m)\geq p$,
which follows from the $p$-robust property.

Next we show that the algorithm generates the random graphs $G(p)$ and $G_2$ in a way  that they satisfy  $G(p)\cup G_2\sim \widetilde G(p)$. We prove it by induction, starting from $i=m$ and progressing downward to $i=1$. For any $i\in [m]$, let $G^i(p)$, $G^i_2$ 
denote the already generated parts of $G(p), G_2$, respectively, after  executing Step 4 \, $m-i+1$ times, so they can only contain edges
with index $\geq i$. Further, let $\widetilde G^i(p)$ be the subgraph
of $\widetilde G(p)$ in which we only keep the edges with index $\geq i$, that is, $\widetilde G^i(p)=\widetilde G(p)-\{e_{i-1},\ldots,e_1\}$.
The inductive proof will show that 
$G^i(p)\cup G^i_2\sim \widetilde G^i(p)$ holds for every $i$.
At the end of the induction, having reached $i=1$, we are going to get  $G^1(p) \cup G^1_2 \sim \widetilde G^1(p)$,
which is the same as $G(p) \cup G_2 \sim \widetilde G(p)$.

Let us consider first the base case  $i=m$. Then we have $\Pr(e_m\in G(p))= \Pr(e_m\in G^m(p))=p$ by Step 4. 
Then in Step 4, edge $e_m$ is put into $G_2$ with probability
$q_m-p'_m$, yielding $\Pr(e_m\in G^m_2)=q_m-p'_m$.
Now observe that the formula (\ref{p'_i}) is chosen such that $p'_i$ is precisely the solution of the equation
\begin{equation}\label{p'eq}
p+ q_i-p'_i -(q_i-p'_i)p \= q_i
\end{equation}
for $p'_i$. 
For $i=m$ the   equation becomes
\begin{equation}\label{p'eqm}
p+ q_m-p'_m  -(q_m-p'_m)p \= q_m,
\end{equation}
and $p'_m=\frac{p(1-q_m)}{1-p}$ is the solution of this equation. Since by Step 4 we have $\Pr(e_m\in G^m(p))=p$ and $\Pr(e_m\in G^m_2)=q_m-p'_m$, therefore,
we get that the left-hand side of (\ref{p'eqm}) is precisely the probability of the event $\{e_m\in G^m(p)\cup G^m_2\}$.
By (\ref{p'eqm}), this probability is equal to $q_m$, which is set to $q_m=\Pr(h_m=1)=\Pr(e_m\in \widetilde G^m(p))$ in Step 2.
This means that $G^m(p)\cup G^m_2\sim \widetilde G^m(p)$, as desired.

For the induction step, assume that the claim is true for $i+1$, i.e., $G^{i+1}(p)\cup G^{i+1}_2\sim \widetilde G^{i+1}(p)$ holds.
In Step 4, edge $e_i$ is added to $G^{i+1}(p)$ with probability $p$. It is also added to $G^{i+1}_2$ with probability 
$q_i-p'_i$.
 Therefore, just like in the base case, we get that 
$p+q_i-p'_i -(q_i-p'_i)p=\Pr(e_i\in G^i(p)\cup G^i_2).$
We already know that $p'_i$ satisfies the equation (\ref{p'eq}), so $e_i$ is added to $\widetilde G^{i+1}(p)$ with probability 
$q_i=\Pr(h_i=1\;|\; h_{i+1}^m=a_{i+1}^m)$, given the already generated part,
represented by $a_{i+1}^m$. By the inductive assumption, $h_{i+1}^m$ is distributed as $\widetilde G^{i+1}(p)$, 
which is the truncated version of $\widetilde G(p)$, keeping only the $\geq i+1$ indexed edges. 
Hence, for  $h_{i+1}^m$, we can write by the chain rule of conditional probabilities:
$$ \Pr(h_{i+1}^m=a_{i+1}^m)\=
\Pr(h_m=a_m)\prod_{j=i+1}^{m-1} \Pr(h_j=a_j\;|\; h_{j+1}^m=a_{j+1}^m).
$$
After processing $e_i$ (i.e., adding it with probability $q_i$), we get 
\begin{eqnarray}\nonumber
\Pr(h_i^m=a_i^m) &=& \Pr(h_i=a_i\;|\; h_{i+1}^m=a_{i+1}^m) \Pr(h_{i+1}^m=a_{i+1}^m) \\ \nonumber
&=& \Pr(h_i=a_i\;|\; h_{i+1}^m=a_{i+1}^m) 
\underbrace{\Pr(h_m=a_m)\prod_{j=i+1}^{m-1} \Pr(h_j=a_j\;|\; h_{j+1}^m=a_{j+1}^m)}_{\Pr(h_{i+1}^m=a_{i+1}^m)}\\ \nonumber
&=& \Pr(h_m=a_m)\prod_{j=i}^{m-1} \Pr(h_j=a_j\;|\; h_{j+1}^m=a_{j+1}^m), 
\end{eqnarray}
which, by the chain rule, is indeed the distribution of $\widetilde G^i(p)$, completing the induction.

Thus, at the end, a realization $a=a_1^m\in \{0,1\}^m$ of $\widetilde G(p)$ is generated with probability 
$$
\Pr(h_1^m=a) \= \Pr(h_m=a_m)\prod_{j=1}^{m-1} \Pr(h_j=a_j\;|\; h_{j+1}^m=a_{j+1}^m),
$$
indeed creating $\widetilde G(p)$ with its correct probability. Therefore,  we get 
$G(p) \cup G_2 \sim \widetilde G(p)$, so $\widetilde G(p)$ arises by adding 
edges to $G(p)$, which cannot destroy a monotone property. This implies the claim of the Lemma.

\hspace*{10mm} \hfill $\spadesuit$

\medskip
Now we have the tools to prove Theorem~\ref{pm}. We repeat its statement below for convenient reference: 
\medskip

\noindent
{\em 
Let $Y=[\xi_{ij}]$ be a random 0--1 matrix of size $n\times n$, drawn from an arbitrary distribution, and let $p\in (0,1)$.
If $Y$ is $p$-robust {\rm (see Definition~\ref{defp-robmx}),} then 
$$
\Pr(Y \, \mbox{\rm has a perfect matching})\geq 1- (n+1)^3 (1-p)^{\lfloor n/2\rfloor}.
$$
}

\medskip
\noindent 
{\bf Proof.}
Let $Z=[\eta_{i,j}]$ be another random 0--1 matrix of size $n\times n$, and choose each entry of $Z$ {\em independently}, 
with $\Pr(\eta_{ij}=1)=p$. If $Z$ is viewed as the bi-adjacency matrix of a bipartite graph, then this will be a random bipartite graph
with bipartition $(V_1,V_2)$, $|V_1|=|V_2|=n$, in which each edge between $V_1, V_2$ is included with probability $p$.
This bipartite random graph model is often denoted by ${\cal G}(n,n,p)$. In our terminology of the $p$-random subgraph
(see Definition~\ref{defp-randsgr}), ${\cal G}(n,n,p)$ is a $p$-random subgraph of the complete bipartite graph $K_{n, n}$.
The bi-adjacency matrix of ${\cal G}(n,n,p)$ is $Z$, where each entry is chosen independently with probability $p$.
Similarly, the $p$-robust random 0--1 matrix $Y$ is the bi-adjacency matrix of a $p$-robust random subgraph (see Definition~\ref{defp-robsgr})
of $K_{n, n}$; let us denote this $p$-robust random subgraph by $\widetilde K_{n, n}(p)$. We assume that the edges are numbered consistently with the matrix precedence relation, see the remark after Definition~\ref{defp-robsgr}.
Then Lemma~\ref{p-robgr} implies that for any monotone graph property $\cal Q$
\begin{equation}\label{mono}
\Pr\lor {\cal G}(n,n,p)\in {\cal Q}\ler \leqq \Pr\lor \widetilde K_{n, n}(p)\in {\cal Q}\ler
\end{equation}
is satisfied. In our case, the property is the existence of a perfect matching, which is a monotone property.

Now let us estimate the probability that the random bipartite graph ${\cal G}(n,n,p)$ has no perfect matching\footnote{This is often analyzed in 
the random graphs literature, see, e.g., Bollob\'as \cite{bollobas}, Janson, Luczak, and Ruci\'nski \cite{janson}, Frieze and Karo\'nski \cite{frieze}, etc.
However, the end-results are typically presented as asymptotic expressions; we could not find a definite reference to a simple, explixit, {\em non-asymptotic} formula for finite graphs, so we derive our own.}.  
By Hall's Theorem\footnote{Originally published in 1935, see Hall \cite{hall}; now it is found in most textbooks on graph theory.}, the graph has no perfect matching if and only if it contains a set $A\subseteq V_1$ or $A\subseteq V_2$, such that $|A|>|\Gamma(A)|$. Let us call such a set $A$ a {\em Hall violator.} 

First we show that if there is a Hall violator
$A$, then there is also one with $|A|\leq \lceil n/2\rceil$. To prove it, let $A_0$ be a minimal Hall violator. If $|A_0|=1$, then we are done,
as $1\leq \lceil n/2\rceil$. If $|A_0|\geq 2$, then for any $x\in A_0$ it holds that $A_0-x$
is already not a Hall violator. Therefore, we must have $|\Gamma(A_0)|=|A_0|-1$, since otherwise we could leave a vertex out of $A_0$, and still 
obtain a Hall violator, contradicting  to the minimality of  $A_0$.

Assume, without loss of generality, that $A_0\subseteq V_1$. If $|A_0| \leq \lceil n/2\rceil$, then we found 
a Hall violator of size $\leq \lceil n/2\rceil$.  Else, if $|A_0|> \lceil n/2\rceil$, then set $B=V_2-\Gamma(A_0)$.
Due to $|\Gamma(A_0)|=|A_0|-1> \lceil n/2\rceil -1$, we have $|\Gamma(A_0)|\geq \lceil n/2\rceil$, so $|B|\leq n-\lceil n/2\rceil\leq \lceil n/2\rceil$.
As $B\cap \Gamma(A_0)=\emptyset$, therefore, $B$ has no neighbor in $A_0$, yielding $\Gamma(B)\cap A_0=\emptyset$.
Then
$$|\Gamma(B)|\leqq n-|A_0| \= n-\underbrace{(|\Gamma(A_0)|+1)}_{|A_0|}\= n-(\underbrace{n-|B|}_{|\Gamma(A_0)|}+1) \leqq |B|-1,
$$
giving $|\Gamma(B)|<|B|$, so $B$ is a Hall violator, too. But we have shown that $|A_0|> \lceil n/2\rceil$ implies $|B|\leq \lceil n/2\rceil$.
Thus, either $A_0$ or $B$ is a Hall violator of size $\leq \lceil n/2\rceil$. This  proves that 
if there is a Hall violator, then there is also one with size $\leq \lceil n/2\rceil$.

Let $M$ denote the event that ${\cal G}(n,n,p)$ has a perfect matching, and its negation is denoted by $\overline{M}$. Further, let $\cal H$ 
be the family of all  Hall violators of size $\leq \lceil n/2\rceil$. We already know that if there is a Hall violator, then there is also one of 
size $\leq \lceil n/2\rceil$, which implies $\overline {M} \= \{{\cal H}\neq \emptyset\}$. Consequently, 
$\Pr(\overline {M}) \= \Pr({\cal H}\neq \emptyset)$.

Let us now estimate $\Pr({\cal H}\neq \emptyset)$. If $A\subseteq V_1$ with $|A|=k$,  then $A\in \cal H$ if and only if 
$1\leq k\leq \lceil n/2\rceil$, and 
$A$ has at least $n-k+1$ non-neighbors in $V_2$. The latter means, there exists a set $B\subseteq V_2$ with  $|B|=n-k+1$, such that there is no edge 
between $A$ and $B$. With the notation $q=1-p$, the probability of having no edge between $A$ and $B$ is $q^{k(n-k+1)}$ (Recall that the edges
in ${\cal G}(n,n,p)$ are independent.) 

To estimate $\Pr({\cal H}\neq \emptyset)$, observe that $A$ can be chosen $n\choose k$ different ways 
from $V_1$, with $1\leq k\leq \lceil n/2\rceil$. For each such $A$, the set $B$ can be chosen in $n \choose n-k+1$ different ways from $V_2$.
Furthermore, we can also symmetrically reverse the positions of $A,B$, so that they trade sides, meaning that   $A\subseteq V_2, B\subseteq V_1$, 
leading to the same probabilities. Thus, we obtain
$$
\Pr(\overline {M}) \= \Pr({\cal H}\neq \emptyset) \leqq 2\sum_{k=1}^{\lceil n/2\rceil} {n\choose k} {n \choose n-k+1} q^{k(n-k+1)}.
$$
To bound the right-hand side from above, we use the symmetry rule ${n \choose \ell} = {n\choose n-\ell}$, and the (crude) upper bound 
${n \choose \ell} \leq n^\ell$. Furthermore, $q^{k(n-k+1)}\leq q^{k(n-k)}$. Then we get
\begin{equation}\label{PrM}
\Pr(\overline {M}) \leqq 2\sum_{k=1}^{\lceil n/2\rceil} {n\choose k} {n \choose k-1} q^{k(n-k)} \leqq
2\sum_{k=1}^{\lceil n/2\rceil} n^{2k}q^{k(n-k)} \=
2\sum_{k=1}^{\lceil n/2\rceil} (n^2q^{n-k})^k.
\end{equation}
Since $k\leq \lceil n/2\rceil$, therefore, $q^{n-k} \leq q^{n-\lceil n/2\rceil}=q^{\lfloor n/2\rfloor}$. Hence, we can continue (\ref{PrM}) as
$$
\Pr(\overline {M}) \leqq 2\sum_{k=1}^{\lceil n/2\rceil} (n^2q^{n-k})^k \leqq 2\sum_{k=1}^{\lceil n/2\rceil} (n^2q^{\lfloor n/2\rfloor})^k.
$$
If $n^2q^{\lfloor n/2\rfloor}< 1$, then the largest summand in the last summation occurs at $k=1$, yielding
\begin{equation}\label{PrM2}
\Pr(\overline {M}) \leqq 2\lceil n/2\rceil n^2q^{\lfloor n/2\rfloor} \leqq (n+1)^3 q^{\lfloor n/2\rfloor},
\end{equation}
which gives us the lower bound 
\begin{equation}\label{CN}
\Pr({M}) \=
\Pr\lor {\cal G}(n,n,p)\; \mbox{has a perfect matching}\ler \geqq 1-(n+1)^3 (1-p)^{\lfloor n/2\rfloor}.
\end{equation}
In this derivation we used the assumption $n^2q^{\lfloor n/2\rfloor}< 1$. However,
if  $n^2q^{\lfloor n/2\rfloor}\geq 1$, then (\ref{PrM2}) still holds, just the upper bound becomes trivial, being larger than 1.
Similarly, (\ref{CN}) becomes trivial with a negative lower bound. Nevertheless, in any case the results (\ref{PrM2}) and (\ref{CN}) provide valid bounds.

Let us use now that having a perfect matching is a monotone property. Then, combining (\ref{CN}) with (\ref{mono}), where the latter is based on
Lemma~\ref{p-robgr}, gives:
\begin{eqnarray}\nonumber
1-(n+1)^3 (1-p)^{\lfloor n/2\rfloor} &\leq& 
\Pr\lor {\cal G}(n,n,p)\; \mbox{has a perfect matching}\ler \\ \label{Knn}
&\leq& \Pr\lor \widetilde K_{n, n}(p)\; \mbox{has a perfect matching}\ler.
\end{eqnarray} 
Recall now that  a perfect matching in  $\widetilde K_{n, n}(p)$ is in 1--1 correspondence with a perfect matching in 
the $p$-robust 0--1 matrix $Y$. (See the remark after Definition~\ref{defp-robsgr}.) Thus, we get from (\ref{Knn}) that 
$$\Pr(Y\; \mbox{has a perfect matching}) \geqq 1-(n+1)^3 (1-p)^{\lfloor n/2\rfloor}$$
holds, completing the proof.

\hspace*{10mm} \hfill $\spadesuit$

Finally, we have all the tools to prove inequality (\ref{pX}), which is needed in the proof of Theorem~\ref{main}.
For convenient reference we repeat here the statement of Theorem~\ref{pm2}:

\medskip
\noindent {\em 
Let $X$ be the $n\times n$ sized random matrix created in {\rm Steps 3} and {\rm 4} of Algorithm 1, and set $n=2N^2$. 
Denote by $p(X)$ the probability that $X$ does not have an $H$-PM.
Then for $N\geq 21$ the value of $p(X)$ is bounded as
$$
p(X)\leq {\rm e}^{-N}.
$$
}

\noindent {\bf Proof.}
Let $B=X^H$ be the $H$-skeleton of $X$ (see Definition~\ref{H-skel}).
 We know from 
Lemma~\ref{p-roblemma} that the following holds:
if $N\geq 4$, then the matrix $B=X^H$ is $p$-robust with $p=\frac{2}{N}$.
Recall that Theorem~\ref{pm} says:
\begin{quote}
{\em Let $Y=[\xi_{ij}]$ be a random 0--1 matrix of size $n\times n$, drawn from an arbitrary distribution, and let $p\in (0,1)$.
If $Y$ is $p$-robust then 
\begin{equation}\label{prpm1}
\Pr(Y \, \mbox{\rm has a perfect matching})\geq 1- (n+1)^3 (1-p)^{\lfloor n/2\rfloor}.
\end{equation}
}
\end{quote}
Combining this with $p=\frac{2}{N}$, and with the fact that $X$ has an $H$-PM if and only if 
$B=X^H$ has a perfect matching, we obtain: if $N\geq 4$, then 
\begin{equation}\label{prpm2}
p(X)\= \Pr(X \, \mbox{\rm has no $H$-PM}) \= \Pr(B \, \mbox{\rm has no perfect matching})\leq (n+1)^3 \left(1-\frac{2}{N}\right)^{\lfloor n/2\rfloor}.
\end{equation}
Substituting $n=2N^2$ yields 
\begin{equation}\label{N2}
p(X)\leqq (2N^2+1)^3 \left(1-\frac{2}{N}\right)^{N^2} \= (2N^2+1)^3 \left(1-\frac{2}{N}\right)^{(N/2)(2N)}
\end{equation}
Let us apply the well-known inequality $1+x\leq {\rm e}^x$ with $x=-2/N$ to get 
$
\left(1-\frac{2}{N}\right) \leqq {\rm e}^{-2/N}.
$
Raising both sides to the power of $N/2$ results in
$
\left(1-\frac{2}{N}\right)^{N/2} \leqq {\rm e}^{-1}.
$
Using this, we can continue (\ref{N2}) as
$$
p(X) \leqq (2N^2+1)^3 \left(1-\frac{2}{N}\right)^{(N/2)(2N)} \leqq (2N^2+1)^3 {\rm e}^{-2N} \=
\underbrace{\left( (2N^2+1)^3 {\rm e}^{-N} \right)}_{A}{\rm e}^{-N}.
$$
We can directly check numerically that for $N\geq 21$ the inequality  $A=(2N^2+1)^3 {\rm e}^{-N}\leq 1$ is satisfied. 
Therefore, for $N\geq 21$ we have 
$$
p(X) \leqq {\rm e}^{-N}.
$$
\hspace*{10mm} \hfill $\spadesuit$

\end{document}